\documentclass[runningheads,envcountsame]{llncs}

\usepackage{environ}
\usepackage{amsfonts,amssymb,caption}
\usepackage{amsmath}
\usepackage{amssymb,latexsym}
\usepackage{extarrows}
\usepackage{mathrsfs}
\usepackage{stmaryrd}
\usepackage[]{algorithm}
\usepackage[noend]{algpseudocode}
\usepackage{mathdots}
\usepackage{tikz}
\usepackage{xcolor}
\usepackage{wrapfig}
\usepackage{comment}
\usepackage{marvosym}

\usepackage{graphicx}

\makeatletter
\def\@citecolor{blue}%
\def\@urlcolor{blue}%
\def\@linkcolor{blue}%

\def\orcidID#1{\smash{\href{http://orcid.org/#1}{\protect\raisebox{-1.25pt}{\protect\includegraphics{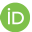}}}}}
\makeatother
\hypersetup{bookmarks,unicode,colorlinks=true}
\usepackage{cleveref}

\usepackage[author=anonymous,margin,footnote,draft]{fixme}
\FXRegisterAuthor{cm}{acm}{CM}
\FXRegisterAuthor{bk}{abk}{BK}
\FXRegisterAuthor{ls}{als}{LS}

\newcommand{\power}{\mathcal P}

\newcommand{\Set}{\mathbf{Set}}

\newcommand{\tbeg}{{\sc T-Beg}}
\renewcommand{\phi}{\varphi}
\newcommand{\upar}{\uparrow\!}

\newcommand{\short}[1]{}
\newcommand{\full}[1]{#1}

\newtoks\prooftoks
\newtoks\temptoks
\def\addproofbody{%
	\temptoks=\expandafter{\BODY}%
	\edef\prooftemp{\the\prooftoks\the\temptoks}%
	\global\prooftoks=\expandafter{\prooftemp}%
}
\NewEnviron{addproof}{%
	\addproofbody
}

\newcommand{\doproofs}{\setcounter{repthms}{0}\the\prooftoks}

\newcounter{repthms}

\def\repthm#1{
	\NewEnviron{rep#1}[1][]{%
		\ifx\hfuzz##1\hfuzz%
		\begin{#1}%
			\else%
			\begin{#1}[##1]%
				\fi%
				\label{repthmx\therepthms}
				\BODY
			\end{#1}%
			\begin{addproof}%
				\noindent\textbf{\Cref{repthmx\therepthms}.\ }
				\begingroup
				\renewcommand{\label}[1]{}
				\itshape%
			\end{addproof}%
			\addproofbody
			\begin{addproof}%
				\endgroup%
				\stepcounter{repthms}
			\end{addproof}%
			\stepcounter{repthms}
		}
	}
	\repthm{theorem}
	\repthm{corollary}
	\repthm{lemma}
	\repthm{proposition}
	
	\NewEnviron{laterproof}{%
		\begin{addproof}\proof{}\end{addproof}%
		\addproofbody%
		\begin{addproof}\endproof\end{addproof}%
	}

\begin{document}

\title{Explaining Non-Bisimilarity in a Coalgebraic Approach: Games
  and Distinguishing Formulas
  \thanks{Work by the first two authors is supported by the DFG project
    BEMEGA (KO 2185/7-2).
    Work by the third author forms part of the DFG project ProbDL2 (SCHR 1118/6-2)}
}
\titlerunning{Explaining Non-Bisimilarity in a Coalgebraic Approach}
%
\author{Barbara K\"{o}nig\inst{1}\orcidID{0000-0002-4193-2889} \and Christina
  Mika-Michalski\inst{1}\Letter \and Lutz Schr\"{o}der\inst{2}\orcidID{0000-0002-3146-5906}}
\authorrunning{B. K\"{o}nig, C. Mika-Michalski and L.
Schr\"{o}der}
%
\institute{University of Duisburg-Essen, Germany \and 
  Friedrich-Alexander-Universit\"at Erlangen-N\"{u}rnberg, Germany 
  \email{\{barbara\_koenig,christina.mika-michalski\}@uni-due.de},
  \email{lutz.schroeder@fau.de}
}
\maketitle              

\begin{abstract}
  Behavioural equivalences can be characterized via bisimulation,
  modal logics, and spoiler-duplicator games. In this paper we work in
  the general setting of coalgebra and focus on generic algorithms for
  computing the winning strategies of both players in a bisimulation
  game. The winning strategy of the spoiler (if it exists) is then
  transformed into a modal formula that distinguishes the given
  non-bisimilar states. The modalities required for the formula are
  also synthesized on-the-fly, and we present a recipe for re-coding
  the formula with different modalities, given by a
  separating set of predicate liftings. Both the game and the
  generation of the distinguishing formulas have been implemented in a
  tool called \tbeg.

\end{abstract}
\section{Introduction}
\label{sec:introduction}

There are many contexts in which it is useful to check whether two
system states are behaviourally equivalent respectively bisimilar. In
this way one can compare a system with its specification, replace a
subsystem by another one that is behaviourally equivalent or minimize
a transition system.
Here we will concentrate on methods for explaining that two given
states in a transition system are \emph{not} bisimilar. The idea is to
provide a witness for non-bisimilarity. Such a witness can
be used to explain (to the user) why an implementation does not
conform to a specification and give further insights for adjusting it.

Two states are bisimilar if they are related by a bisimulation
relation. But this definition does not provide us with an immediate
witness for non-bisimilarity, since we would have to enumerate all
relations including that particular pair of states and show that they
are not bisimulations. Hence, we have to resort to other
characterizations of bisimilarity: bisimulation games
\cite{Stirling_bisim_games}, also known as spoiler-duplicator games,
and modal logic. In the former case a proof of the non-bisimilarity of
two states is given by a winning strategy of the spoiler. In the
latter case the Hennessy-Milner theorem
\cite{HennessyMilner_Logic:1985} guarantees for image-finite labelled
transition systems that, given two non-bisimilar states $x_0,x_1$, there
exists a modal formula $\phi$ such that one of the states
satisfies $\phi$ and the other does not.  The computation of such
distinguishing formulas is explained
in~\cite{c:automatically-explaining-bisim}.

While the results and techniques above have been introduced for
labelled transition systems, we are here interested in the more
general setting of coalgebras \cite{Rut03:universal}, which encompass
various types of transition systems. Here we concentrate on coalgebras
living in $\Set$, where an endofunctor $F\colon\Set\to\Set$ specifies
the branching type of the coalgebra (non-deterministic, probabilistic,
etc.).

Modal logics have been extensively studied for coalgebras and it has
been shown that under certain restrictions, modal coalgebraic logic is
expressive, i.e., it satisfies the Hennessy-Milner theorem
\cite{PATTINSON2003177,SCHRODER2008230}. However, to our knowledge, no
explicit construction of distinguishing formulas in the coalgebraic
setting has yet been given.

Coalgebraic games have been studied to a lesser extent: we
  refer to Baltag \cite{clg:Baltag,BALTAG200042}, where the game is based on
  providing subsets of bisimulation relations (under the assumption that
  the functor $F$ is weak pullback preserving) and a generalization of
  Baltag's game to other functors in
  \cite{k:terminal-sequence-games}. Furthermore there is our own
contribution \cite{km:bisim-games-logics-metric}, on which this
article is based, and \cite{kkhkh:codensity-games}, which considers
codensity games from an abstract, fibrational perspective.

We combine both the game and the modal logic view on
  coalgebras and present the following contributions:

\smallskip

\noindent$\triangleright$ \emph{We describe how to compute
    the winning strategies of the players in the behavioural
    equivalence game.}

\noindent$\triangleright$ \emph{We show how to construct a
    distinguishing formula based on the spoiler strategy. The
    modalities for the formula are not provided a priori, but are
    synthesized on-the-fly as so-called \emph{cone modalities} while
    generating the formula.}

\noindent$\triangleright$ \emph{Finally we show under which
    conditions one can re-code a formula with such modalities into a
    formula with different modalities, given by a separating set of
    predicate liftings.}

\smallskip

Both the game and the generation of the distinguishing formulas have
been implemented in a generic tool called \tbeg \footnote{Available
  at:
  \texttt{https://www.uni-due.de/theoinf/research/tools\_tbeg.php}},
where the functor is provided as a parameter. In particular, using
this tool, one can visualize coalgebras, play the game (against the
computer), derive winning strategies and convert the winning strategy
of the spoiler into a distinguishing formula.
Since the development of the tool was our central aim, we have made
design decisions in such a way that we obtain effective
algorithms. This means that we have taken a hands-on approach and
avoided constructions that potentially iterate over infinitely many
elements (such as the set of all modalities, which might be infinite).
The partition refinement algorithm presented in the paper distinguishes states that are not behaviourally
equivalent by a single equivalence class compared to other
  techniques which iterate over the final chain
  \cite{kk:coalgebra-weighted-automata-journal,DorschEA17}. Separation
  via a single equivalence class is a technique used within known
algorithms for checking bisimilarity in labelled transition systems
\cite{Kanellakis1990CCSEF,DBLP:journals/siamcomp/PaigeT87}. This
requires a certain assumption on the endofunctor specifying the
branching type (dubbed \emph{separability by singletons}). Note that
\cite{DBLP:journals/siamcomp/PaigeT87} has already been generalized to
a coalgebraic setting in \cite{DorschEA17}, using the assumption of
\emph{zippability}. Here we compare these two assumptions.

After presenting the preliminaries (Section~\ref{sec:preliminaries}),
including the game, we describe how to compute the winning strategies
in Section~\ref{sec:gen_strategies}. In Section~\ref{sec:gen_formula}
we show how to construct and re-code distinguishing formulas, followed
by a presentation of the tool \tbeg\ in
Section~\ref{sec:tbeg}. Finally, we conclude in
Section~\ref{sec:conclusion}. \full{The proofs can be found in
  Appendix~\ref{sec:proofs}.}\short{The proofs can be found in the
  full version of the paper
  \cite{kms:non-bisimilarity-coalgebraic-arxiv}.}

\section{Preliminaries}
\label{sec:preliminaries}

\paragraph*{Equivalence relations and characteristic functions:}

Let $R \subseteq X \times X$ be an \emph{equivalence relation}, where
the set of all equivalence relations on $X$ is given by
$\mathit{Eq}(X)$. For $x_0\in X$ we denote the \emph{equivalence class}
of $x_0$ by $[x_0]_R = \{x_1 \in X \mid (x_0,x_1) \in R\}$. By $E(R)$ we denote
the set of all equivalence classes of $R$. Given $Y\subseteq X$, we
define the \emph{$R$-closure} of $Y$ as follows:
$[Y]_R=\{x_1 \in X \mid \exists \, x_0 \in Y\, (x_0,x_1) \in R \}$.

For $Y\subseteq X$, we denote its \emph{predicate} or
\emph{characteristic function} by $\chi_Y\colon X\to
\{0,1\}$. Furthermore, given a characteristic function
$\chi\colon X\to \{0,1\}$, its corresponding set is denoted
$\hat{\chi}\subseteq X$.

We will sometimes overload the notation and for instance write $[p]_R$
for the $R$-closure of a predicate $p$. Furthermore we will write
$p_0\cap p_1$ for the intersection of two predicates.

\paragraph*{Coalgebra:}

We restrict our setting to the category $\Set$, in particular we
assume an \emph{endofunctor} $F\colon \Set\to \Set$, intuitively
describing the branching type of the transition system under
consideration.
A \emph{coalgebra} \cite{Rut03:universal}, describing a transition
system of this branching type, is given by a function
$\alpha\colon X\to FX$.  Two states $x_0,x_1\in X$ are \emph{behaviourally
  equivalent} ($x_0\sim x_1$) if there exists a coalgebra homomorphism $f$
from $\alpha$ to some coalgebra $\beta\colon Y\to FY$ (i.e., a
function $f\colon X\to Y$ with $\beta\circ f = Ff\circ \alpha$) such
that $f(x_0) = f(x_1)$.  We assume that $F$ preserves weak pullbacks,
which means that behavioural equivalence and coalgebraic bisimilarity
  coincide, and we will use the two terms interchangeably.

\paragraph*{Preorder lifting:}
Furthermore we need to \emph{lift} preorders under a functor $F$. To
this end, we use the lifting introduced in
\cite{bk:finitary-functors-set-preord-poset} (essentially the standard
\emph{Barr extension} of~$F$~\cite{Barr70,Trnkova80}), which
guarantees that the lifted relation is again a preorder provided
that~$F$ preserves weak pullbacks: Let $ \leq $ be a preorder on $Y$,
i.e.  $ \leq \ \subseteq Y \times Y $. We define a preorder $ \leq^F$
on $FY$ by $ t_0 \leq^F t_1 $ iff there exists $ t \in F(\leq) $ such
that $ F\pi_i(t) = t_i$ for $i \in \{0,1\} $, where
$ \pi_i \colon \leq\ \to Y$ are the usual projections. More
concretely, we consider the order $\leq \ = \{(0,0), (0,1), (1,1)\}$
over $2=\{0,1\}$ and its corresponding liftings $\leq^F$.

Note that applying the functor is monotone wrt. the lifted order:

\begin{lemma}[\cite{km:bisim-games-logics-metric}]
  \label{lem:lift-order-pred}
  Let $(Y,\le)$ be an ordered set and let $p_0,p_1\colon X\to Y$ be
  functions. Then $ p_{0} \leq p_{1} $ implies
  $Fp_{0} \leq^{F} Fp_{1} $, with both inequalities read pointwise.
\end{lemma}

\paragraph*{Predicate liftings:}

In order to define the modal logic, we need the notion of
\emph{predicate liftings} (also called \emph{modalities}). Formally, a
predicate lifting for $F$ is a natural transformation
$\bar \lambda\colon\mathcal{Q} \Rightarrow \mathcal{Q}F$, where
$\mathcal{Q}$ is the contravariant powerset functor. It transforms
subsets $P\subseteq X$ into subsets $\bar \lambda(P) \subseteq FX$.

We use the fact that predicate liftings are in one-to-one
correspondence with functions of type $\lambda\colon F2\to 2$ (which
specify subsets of $F2$ and will also be called \emph{evaluation
  maps}) \cite{SCHRODER2008230}. We view subsets $P\subseteq X$ as
predicates $p = \chi_P$ and lift them via $p\mapsto \lambda\circ Fp$.
In order to obtain expressive logics, we also need the notion of a
separating set of predicate liftings.

\begin{definition}
  \label{def:separating-functor}
  A set $\Lambda$ of evaluation maps for a functor
  $F\colon \Set\to \Set$ is \emph{separating} if for all sets $X$ and
  $t_0,t_1\in FX$ with $t_0 \neq t_1$, there exists
  $\lambda\in \Lambda$ and $p\colon X\to 2$ such that
  $\lambda(Fp(t_0)) \neq \lambda(Fp(t_1))$.
\end{definition}
\noindent This means that every $t \in FX$ is uniquely determined by
the set
$ \{(\lambda, p) \mid \lambda\in\Lambda, p\colon X\to 2,
\lambda(Fp(t)) = 1 \} $. Such a separating set of predicate liftings
exists iff $(Fp \colon FX \rightarrow F2)_{p\colon X \rightarrow 2}$
is jointly injective.

Here we concentrate on \emph{unary} predicate liftings: If one
generalizes to \emph{polyadic} predicate liftings, a separating set of
predicate liftings can be found for every accessible functor
\cite{SCHRODER2008230}.

Separating sets of monotone predicate liftings and the lifted order on
$F2$ are related as follows:

\begin{proposition}[\cite{km:bisim-games-logics-metric}]
  \label{prop:ev-monotone}
  An evaluation map $\lambda\colon F2\to 2$ corresponds to a monotone
  predicate lifting
  $(p\colon X\to 2)\mapsto (\lambda\circ Fp\colon FX\to 2)$ iff
  $\lambda\colon (F2,\le^F)\to (2,\le)$ is monotone.
\end{proposition}

\begin{proposition}[\cite{km:bisim-games-logics-metric}]
   \label{prop:separating-pl-antisym-ji}
   $F$ has a separating set of monotone predicate liftings iff
   $\leq^F\subseteq F2\times F2$ is anti-symmetric and
   $(Fp \colon FX \rightarrow F2)_{p\colon X \rightarrow 2}$ is
   jointly injective.
 \end{proposition}

\paragraph*{Coalgebraic modal logics:}

Given a cardinal $\kappa$ and a set $\Lambda$ of evaluation maps
$\lambda\colon F2\to 2$, we define a coalgebraic modal language
$\mathcal{L}^{\kappa}(\Lambda)$ via the grammar
\[ \phi ::= \bigwedge \Phi \mid \neg \phi \mid [\lambda] \phi
\quad\text{where $\Phi \subseteq \mathcal{L}^{\kappa}(\Lambda)$ with
  $\mathit{card}(\Phi) < \kappa$ and $\lambda \in \Lambda $.} \]
The last case describes the prefixing of a formula $\phi$ with a
modality $[\lambda]$. Given a coalgebra $\alpha\colon X\to FX$ and a
formula $\phi$, the semantics of such a formula is given by a map
$\llbracket\phi\rrbracket_\alpha\colon X\to 2$, where conjunction and
negation are interpreted as usual and
$\llbracket [\lambda]\phi \rrbracket_{\alpha}=\lambda\circ
F\llbracket\phi\rrbracket_\alpha\circ \alpha$.

For simplicity we will often write $\llbracket \phi\rrbracket$ instead
of $\llbracket \phi\rrbracket_\alpha$. Furthermore for $x\in X$, we
write $x\models \phi$ whenever $\llbracket \phi \rrbracket(x) = 1$. As
usual, whenever
$\llbracket \phi\rrbracket_\alpha = \llbracket \psi\rrbracket_\alpha$
for all coalgebras $\alpha$ we write $\phi\equiv \psi$.
We will use derived operators such as $\mathit{tt}$ (empty
conjunction), $\mathit{ff}$ ($\lnot \mathit{tt}$) and $\bigvee$
(disjunction).

The logic is always adequate, i.e., two
behaviourally equivalent states satisfy the same formulas. Furthermore
whenever $F$ is $\kappa$-accessible and the set $\Lambda$ of predicate
liftings is separating, it can be shown that the logic is also
expressive, i.e., two states that satisfy the same formulas are
behaviourally equivalent~\cite{Pattinson04,SCHRODER2008230}.

\paragraph*{Bisimulation game:}
We will present the game rules first introduced in
\cite{km:bisim-games-logics-metric}. At the beginning of a game, two
states $x_0,x_1$ are given. The aim of the spoiler (S) is to prove that
$x_0 \nsim x_1$, the duplicator (D) attempts to show $x_0 \sim x_1$.

\begin{itemize}\label{def:game}
  \itemsep0pt
\item \textbf{Initial configuration:} A coalgebra
  $ \alpha \colon X \to FX $ and a position given as pair
  $(x_0,x_1) \in X \times X$. From a position $(x_0,x_1)$, the game
  play proceeds as follows:
\item \textbf{Step~1:} $S$ chooses $j\in \{0,1\}$, (i.e.\ $x_0$ or
  $x_1)$ , and a predicate $p_j\colon X \to 2$.
\item \textbf{Step~2:} $D$ must respond for $x_{1-j}$ with a predicate
  $p_{1-j}$ satisfying
  \[ Fp_{j}(\alpha(x_j)) \leq^{F} Fp_{1-j}(\alpha(x_{1-j})). \]
\item \textbf{Step~3:} $S$ chooses $\ell \in \{0,1\}$ (i.e.\ $p_0$ or
  $p_1$) and an $x'_\ell \in X$ with $ p_\ell(x'_\ell)=1$.
\item \textbf{Step~4:} $D$ must respond with an $x'_{1-\ell} \in X$
  such that $ p_{1-\ell}(x_{1-\ell}')=1$.
\end{itemize}

\noindent After one round the game continues in Step~$1$ with the pair
$ (x_0',x_1')$. $D$ wins if the game continues forever or if $S$ has
no move at Step~$3$. In all other cases, i.e. $D$ has no move at
Step~$2$ or Step~$4$, $S$ wins.

\smallskip

This game generalizes a bisimulation game for probabilistic transition
systems from \cite{dlt:approx-analysis-prob}. Note that -- different
from the presentation in \cite{dlt:approx-analysis-prob} -- we could
also restrict the game in such a way that $S$ has to choose index
$\ell=1-j$ in Step~3.

We now give an example that illustrates the differences between our generic game and the classical bisimulation game for labelled transition systems
\cite{Stirling_bisim_games}.
\begin{example}
	\label{ex:classical-game}
	Consider the transition system in Figure \ref{fig:nondet-ts}, which
	depicts a coalgebra $\alpha\colon X\to FX$, where
	$F = \mathcal{P}_f(A\times (-))$ specifies 
	finitely branching labelled transition systems. Clearly $x_0 \nsim x_1$.

	{\makeatletter
		\let\par\@@par
		\par\parshape0
		\everypar{}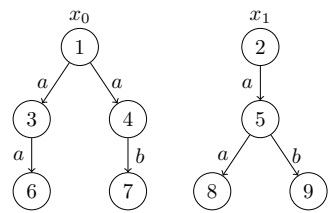
\begin{wrapfigure}{r}{0.4\textwidth}	
			\centering
			\vspace{-10pt}
			\scalebox{0.8}{	
				\begin{tikzpicture}
				\node (S) at (0,-1.5) {$x_0$}; 			
				\node (S1) at (0,-2) [circle,draw]{$1$}; 			
				\node (S3) at (-0.8,-3.2) [circle,draw]{$3$};
				\node (S4) at (0.8,-3.2) [circle,draw]{$4$}; 
				\draw  [->] (S1) to node [left]{$a$} (S3);
				\draw  [->] (S1) to node [right]{$a$} (S4);	
				\node (S) at (3,-1.5) {$x_1$}; 			
				\node (T2) at (3,-2) [circle,draw]{$2$}; 			
				\node (T5) at (3,-3.2) [circle,draw]{$5$};  
				\node (S6) at (-0.8,-4.4) [circle,draw]{$6$}; 			
				\node (S7) at (0.8,-4.4) [circle,draw]{$7$};  
				\node (T8) at (2.2,-4.4) [circle,draw]{$8$}; 			
				\node (T9) at (3.8,-4.4) [circle,draw]{$9$};  
				\draw  [->] (T2) to node [left]{$a$} (T5);
				\draw  [->] (S3) to node [left]{$a$} (S6);
				\draw  [->] (S4) to node [right]{$b$} (S7);
				\draw  [->] (T5) to node [left]{$a$} (T8);
				\draw  [->] (T5) to node [right]{$b$} (T9);
				\end{tikzpicture}	}		 			
			\caption{Spoiler has a winning strategy at $(x_0,x_1)$.}
			\label{fig:nondet-ts}		
		\end{wrapfigure}First consider the classical game where one possible winning
		strategy of the spoiler is as follows: he moves
		$x_0 = 1 \stackrel{a}{\to} 4$, which must be answered by the
		duplicator via $x_1 = 2\stackrel{a}{\to} 5$. Now the spoiler switches
		and makes a move $5\stackrel{a}{\to} 8$, which can not be answered
		by the duplicator. 
		
		In our case a corresponding game proceeds as follows: the spoiler chooses $j=0$ and $p_0 = \chi_{\{4\}}$. Now the
		duplicator takes $x_{1}$ and can for instance answer with $p_1 =
		\chi_{\{5\}}$, which leads to
		\par}		  
              \[ Fp_0(\alpha(x_0)) = \{(a,0),(a,1)\} \le^F \{(a,1)\} =
                Fp_1(\alpha(x_1)) \] (Compare this with the
              visualization of the order $\le^F$ on $F2$ in
              Figure~\ref{fig:cone_modalities_pow}.) Regardless of how
              $S$ and $D$ choose states, the next game configuration is
              $(4,5)$.
	
	Now the spoiler is \emph{not} forced to switch, but can
	choose $j=0 $ (i.e. $4$) and can play basically any predicate $p_0$, which leads
	to either $Fp_0(\alpha(4)) = \{(b,1)\}$ or
	$Fp_0(\alpha(4)) = \{(b,0)\}$.  $D$ has no answering move, since
	$Fp_1(\alpha(5))$ will always contain tuples with $a$ \emph{and}
	$b$, which are not in $\le^F$-relation with the move of $S$ (see
	also Figure~\ref{fig:cone_modalities_pow}, which depicts $F2$ and
	its order).
\end{example}
\noindent The game characterizes bisimulation for functors that are
weak pullback preserving and for which the lifted order $\le^F$ is
anti-symmetric. Then it holds that $x_0\sim x_1$ if and only if $D$ has a
winning strategy from the initial configuration $(x_0,x_1)$.  As already shown
in \cite{km:bisim-games-logics-metric}, in the case of two
non-bisimilar states $x_0\nsim x_1$ we can convert a modal formula
$\phi$ distinguishing $x_0,x_1$, i.e., $x_0\models \phi$ and
$x_1\not\models \phi$, into a winning strategy for the
spoiler. Furthermore we can extract the winning strategy for the
duplicator from the bisimulation relation.

However, in \cite{km:bisim-games-logics-metric} we did not yet show how to
directly derive the winning strategy of both players or how to
construct a distinguishing formula $\phi$.

\section{Computation of Winning Strategies}
\label{sec:gen_strategies}

In the rest of the paper we will fix a coalgebra
$\alpha:X \rightarrow FX$ with finite $X$ for a weak pullback
preserving endofunctor $F\colon \Set\to\Set$. Furthermore we assume
that $F$ has a separating set of \emph{monotone} predicate liftings,
which implies that $\le^F$, the lifted order on $2$, is
anti-symmetric, hence a partial order.

We first present a simple but generic partition refinement algorithm to derive the winning strategy for the spoiler ($S$) and duplicator ($D$) for a given
coalgebra $\alpha\colon X \to FX$. This is based on a fixpoint
iteration that determines those pairs of states $(x_0,x_1) \in X \times X$
for which $D$ has a winning strategy, i.e. $x_0\sim x_1$.
  In particular we consider the relation $W_{\alpha}$, which -- as we
  will show --- is the greatest fixpoint of the following monotone
  function
  $\mathcal{F}_{\alpha}\colon \mathit{Eq}(X) \to \mathit{Eq}(X)$ on
  equivalence relations:
\begin{eqnarray*}
  \mathcal{F}_{\alpha}(R) & = & \{(x_0,x_1) \in R \mid \forall P\in E(R)\colon
  F\chi_P(\alpha(x_0)) = F\chi_P(\alpha(x_1)) \} \\
  W_{\alpha} & = & \{(x_0,x_1) \in X \times X \mid \text{there exists a winning
  	strategy of $D$ for } (x_0,x_1) \} 
\end{eqnarray*}

\begin{theorem}[\cite{km:bisim-games-logics-metric}]
	\label{thm:winning-strategy-defender}
	Assume that $F$ preserves weak pullbacks and has a separating set of
	monotone evaluation maps. Then $ x_0 \sim x_1 $ iff D has a
	winning strategy for the initial configuration $ (x_0,x_1) $.
\end{theorem}

\noindent  In the following, we will prove that the greatest fixpoint of
$\mathcal{F}_{\alpha}$ (i.e.\ $\nu \mathcal{F}_{\alpha}$) coincides
with $W_{\alpha}$ and hence gives us bisimilarity. Note that
  $\mathcal{F}_\alpha$ splits classes with respect to only a single
  equivalence class $P$. This is different from other coalgebraic
  partition refinement algorithms where the current equivalence
  relation is represented by a surjection $e$ with domain $X$ and we
  separate $x_0,x_1$ whenever $Fe(\alpha(x_0)) \neq Fe(\alpha(x_1))$,
  which intuitively means that we split with respect to all
  equivalence classes at once.  Hence we will need to impose extra
requirements on the functor, spelled out below, in order to obtain
this result.

One direction of the proof deals with deriving a winning strategy for
$S$ for each pair $(x_0,x_1) \notin \nu \mathcal{F}_{\alpha} $. In order
to explicitly extract such a winning strategy for $S$ -- which will
also be important later when we construct the distinguishing formula
-- we will slightly adapt the algorithm based on fixpoint
iteration. Before we come to this, we formally define and explain the
strategies of $D$ and $S$.

We start with the winning strategy of the duplicator in the case where
the two given states are bisimilar. This strategy has already been
presented in \cite{km:bisim-games-logics-metric}, but we describe it here
again explicitly. The duplicator only has to know a suitable coalgebra
homomorphism. 

\begin{proposition}[Strategy of the duplicator,
  \cite{km:bisim-games-logics-metric}]
  Let $\alpha\colon X\to FX$ be a coalgebra. Assume that $D$, $S$ play
  the game on an initial configuration $(x_0,x_1)$ with $x_0\sim x_1$. This means
  that there exists a coalgebra homomorphism $f\colon X\to Z$ from
  $\alpha$ to a coalgebra $\beta\colon Z\to FZ$ such that $f(x_0)=f(x_1)$.

  Assume that in Step~2 $D$ answers with
  $p_{1-j} = [p_j]_{\mathit{ker}(f)}$, i.e., $p_{1-j}$ is the
  $\mathit{ker}(f)$-closure\footnote{For a function $f\colon X\to Y$,
    $\mathit{ker}(f) = \{(x_0,x_1)\mid x_0,x_1\in X, f(x_0) = f(x_1)\}
    \subseteq X\times X$.} of the predicate $p_j$. (In other words:
  $p_{1-j}(s) = 1$ iff there exists $t\in X$ such that $f(s) = f(t)$ and
  $p_j(t) = 1$.)

  Then the condition of Step~2 is satisfied and in Step~4 $D$ is
  always able to pick a state $x_{1-\ell}'$ with
  $p_{1-\ell}(x_{1-\ell}') = 1$ and $f(x_\ell')=f(x_{1-\ell}')$.
\end{proposition}

\noindent We argue why this strategy is actually winning: Since $f$ is
a coalgebra homomorphism we have
$Ff(\alpha(x_0)) = \beta(f(x_0)) = \beta(f(x_1)) =
Ff(\alpha(x_1))$. By construction, $p_{1-j}$ factors through $f$, that
is $p_{1-j} = p'_j\circ f$ for some $p'_j\colon Z\to 2$. This implies
$Fp_{1-j}(\alpha(x_j)) = Fp_j'(Ff(\alpha(x_j))) =
Fp_j'(Ff(\alpha(x_{1-j}))) \allowbreak =
Fp_{1-j}(\alpha(x_{1-j}))$. Since $p_j\le p_{1-j}$ it follows from
monotonicity (Lemma~\ref{lem:lift-order-pred}) that
$Fp_j(\alpha(x_j)) \le^F Fp_{1-j}(\alpha(x_j)) =
Fp_{1-j}(\alpha(x_{1-j}))$.  Hence $p_{1-j}$ satisfies the conditions
of Step~2. Furthermore if the spoiler picks a state $x_\ell'$ in $p_j$
in Step~3, the duplicator can pick the same state in $p_{1-j}$ in
Step~4. If instead the spoiler picks a state $x_{\ell}'$ in $p_{1-j}$,
the duplicator can, due to the closure, at least pick a state
$x_{1-\ell}'$ in $p_j$ which satisfies $f(x_{1-\ell}') = f(x_\ell')$,
which means that the game can continue.

\smallskip

We now switch to the spoiler strategy that can be used to explain why
the states are not bisimilar. A strategy for the spoiler is given by a pair of functions

      \[ I\colon X \times X \rightarrow \mathbb{N}_0\cup\{\infty\} \text{
      	and } T\colon (X\times X)\backslash \nu \mathcal{F}_\alpha
      \rightarrow X \times \power{X}. \]
      
    \noindent Here, $I(x_0,x_1)$ denotes the first index where
    $x_0,x_1$ are separated in the fixpoint iteration of
    $\mathcal{F}_\alpha$.  The second component $T$ tells the spoiler
    what to play in Step~1. In particular whenever
    $T(x_0,x_1) = (x_j,P)$, $S$ will play $j$ (uniquely determined by
    $x_j$ unless $x_0=x_1$, in which case~$S$ does not win) and
    $p_j = \chi_{P}$.

In the case $I(x_0,x_1) < \infty$ such a winning strategy for $S$ can be computed during fixpoint
iteration, see Algorithm~\ref{AlgorithmA_extended}.  Assume that the
algorithm terminates after $n$ steps and returns $R_n$. It is easy to
see that $R_n$ coincides with $\nu \mathcal{F}_\alpha$: as usual for 
partition refinement, we start with the coarsest relation
$R_0 = X\times X$. Since $\le^F$ is, by assumption, anti-symmetric
$ F\chi_{P}(\alpha(x_0)) \leq^{F} F\chi_{P}(\alpha(x_1))$ and
$ F\chi_{P}(\alpha(x_1)) \leq^{F} F\chi_{P}(\alpha(x_0))$ are
equivalent to $ F\chi_{P}(\alpha(x_0)) = F\chi_{P}(\alpha(x_1))$ and
the algorithm removes a pair $(x_0,x_1)$ from the relation iff this
condition does not hold. In addition, $ T (x_0, x_1) $ and $ I (x_0, x_1) $ are updated, where we distinguish whether $Fp (\alpha(x_0)) \nleq^F Fp (\alpha(x_1))$ or $Fp (\alpha(x_0)) \ngeq^F Fp (\alpha(x_1))$ hold.

Every relation $R_i$ is finer than its predecessor $R_{i-1}$ and,
since $\mathcal{F}_\alpha$ preserves equivalences, each~$R_i$ is an
equivalence relation. Since we are assuming a finite set $X$ of
states, the algorithm will eventually terminate.

\begin{algorithm}[h]
  \begin{algorithmic}[1]
    \Procedure{Compute greatest fixpoint of $\mathcal{F}_{\alpha}$ and winning moves for $S$}{}
    \ForAll { $(x_0,x_1) \in X \times X $} 	\State $I(x_0,x_1) \gets \infty$ \EndFor
    \State $i \gets 0$, $R_{0} \gets X \times X$
    \Repeat 
    \State $i \gets i+1$, $R_i \gets R_{i-1}$
    \ForAll {$(x_0,x_1) \in R_{i-1}$} 		
    \ForAll { $P \in E(R_{i-1})$} 	
    \If {$ F\chi_P(\alpha(x_0)) \nleq^{F} F\chi_P(\alpha(x_1))$}  
    \State $T(x_0,x_1) \gets (x_0,P)$, $I(x_0,x_1) \gets i$, $R_{i} \gets R_{i} \setminus \{(x_0,x_1)\}$
    \Else \If {$ F\chi_P(\alpha(x_1)) \nleq^{F} F\chi_P(\alpha(x_0))$}  
    \State $T(x_0,x_1) \gets (x_1,P)$, $I(x_0,x_1) \gets i$, $R_{i} \gets R_{i} \setminus \{(x_0,x_1)\}$
    \EndIf
    \EndIf
    \EndFor
    \EndFor
    \Until{$R_{i-1}= R_i$} \\
    \Return $R_i,T,I$	
    \EndProcedure
  \end{algorithmic}
  \caption{Computation of $\nu \mathcal{F}_\alpha$ and the winning
    strategy of the spoiler} \label{AlgorithmA_extended}
\end{algorithm}

We will now show that Algorithm~\ref{AlgorithmA_extended} indeed computes a winning strategy
for the spoiler.

\begin{repproposition}
  \label{prop:constr-winning-str}
  Assume that $R_n = \nu \mathcal{F}_\alpha, T,I$ have been computed
  by Algorithm~\ref{AlgorithmA_extended}. Furthermore let
  $(x_0,x_1) \notin R_n$, which means that $I(x_0,x_1) < \infty$ and $T(x_0,x_1)$
  is defined. Then the following constitutes a winning strategy for
  the spoiler:
  \begin{itemize}
  \item Let $T(x_0,x_1) = (x_j,P)$. Then in Step~1 $S$ plays
    $j \in \{0,1\}$ and the predicate $p_j=\chi_{P}$.
  \item Assume that in Step~2, $D$ answers with a state $x_{1-j}$ and a
    predicate $p_{1-j}$ such that
    $ Fp_{j}(\alpha(x_j)) \leq^{F} Fp_{1-j}(\alpha(x_{1-j})) $.
  \item Then, in Step~3 there exists a state $x_{1-j}'\in X$ such
    that $p_{1-j}(x_{1-j}')=1$ and
    $I(x_j',x_{1-j}') < I(x_0,x_1)$ for all $x_j'\in X$ with
    $p_j(x_j')=1$. $S$ will hence select $\ell=1-j$, i.e. $p_{1-j}$, and this state
    $x_{1-j}'$.
  \item In Step~4, $D$ selects some $x_j'$ with $p_{j}(x_j')=1$ and
    the game continues with $(x_0',x_1')$ where $(x_0',x_1')\in R_n$
    and $I(x_0',x_1') < I(x_0,x_1)$.
  \end{itemize}
\end{repproposition}

\begin{laterproof}~
	We have to show that whenever we reach Step~3 there always exists a
	state $x_{1-j}'\in X$ such that $p_{1-j}(x_{1-j}') = 1$ and $I(x_j',x_{1-j}') < I(x_0,x_1) $ for
	all $x_j'\in X$ with $p_j(x_j') = 1$.
	
	Let us first observe that $p_{1-j}\nleq p_j$. If this were the case, we
	would have
	$Fp_j (\alpha(x_j)) \leq^F Fp_{1-j}(\alpha(x_{1-j})) \leq^F Fp_j(\alpha(x_{1-j}))$.
	But $\{x_0,x_1\} = \{x_j,x_{1-j}\}$ are separated at Step~$I(x_0,x_1)=i$ precisely
	because this inequality does not hold for $p_j$ which represents one
	of the equivalence classes of $R_{i-1}$.
	
	Hence there exists an $x_{1-j}'\in X$ such that $p_{1-j}(x_{1-j}') = 1$ and
	$p_j(x_{1-j}') = 0$.
	
	Since the equivalence relations $R_i$ are subsequently refined by
	the algorithm, $p_j$ -- being an equivalence class of $R_{i-1}$ --
	is a union of equivalence classes of $R_n$. So, since $x_{1-j}'$ is not
	contained in $P = \hat{p_j}$, it is not in $R_{i-1}$-relation to
	any $x_j'\in P$, hence $I(x_j',x_{1-j}') \le i-1$ for all such $x_j'$.
	
	Since the index $I(x_0,x_1)$ decreases after every round of the game,
	$D$ will eventually not be able to find a suitable answer in Step~2
	and will lose.
	\qed
\end{laterproof}

\noindent Finally, we show that $\nu \mathcal{F}_{\alpha}$ coincides with
$W_{\alpha}$ and therefore also with behavioural equivalence $\sim$
(see \cite{km:bisim-games-logics-metric}). For this purpose, we need one further
requirement on the functor:
 
\begin{definition}
  Let $F\colon\Set\to\Set$ be an endofunctor on $\Set$. Given a
  set~$X$,~$F$ is \emph{separable by singletons on~$X$} if the
  following holds: for all $t_0 \neq t_1$ with $t_0,t_1\in FX$, there
  exists $p\colon X\to 2$ where $p(x) = 1$ for exactly one $x\in X$
  (i.e., $p$ is a singleton) and $Fp(t_0) \neq Fp(t_1)$.
  Moreover,~$F$ is \emph{separable by singletons} if~$F$ is separable
  by singletons on all sets $X$.
\end{definition}
\noindent
It is obvious that separability by singletons implies the existence of
a separating set of predicate liftings, however the reverse implication
does not hold as the following example shows.

\begin{example}
  \label{ex:separable-singletons}

  A functor that does not have this property, but does have a
  separating set of predicate liftings, is the monotone neighbourhood
  functor $\mathcal{M}$ with
  $\mathcal{M}X = \{Y\in \mathcal{Q}\mathcal{Q}X\mid \text{$Y$
    upwards-closed}\}$ (see e.g. \cite{HansenKupke04}), where
  $\mathcal{Q}$ is the contravariant powerset functor. Consider
  $X = \{a,b,c,d\}$ and $t_0,t_1\in\mathcal{M}X$ where
  $t_0 =\ \upar\{\{a,b\},\{c,d\}\}$,
  $t_1 =\ \upar\{\{a,b,c\},\{a,b,d\},\{c,d\}\}$. That is, the only
  difference is that $t_0$ contains the two-element set $\{a,b\}$ and
  $t_1$ does not. For any singleton predicate~$p$, the image of
  $\mathcal{Q}p\colon \mathcal{P}2\to \mathcal{P}X$ does not contain a
  two-element set, hence $\mathcal{M}p(t_0) = \mathcal{M}p(t_1)$ --
  since~$t_1$ and~$t_2$ agree on subsets of~$X$ of cardinality
  different from~$2$ -- and $t_0,t_1$ cannot be distinguished.

  By contrast, both the finite powerset functor $\mathcal{P}_f$ and
  the finitely supported probability distribution functor
  $\mathcal{D}$ (which are both $\omega$-accessible and hence yield a
  logic with only finite formulas) are separable by singletons.
\end{example}
\noindent As announced, separability by singletons implies that the
fixpoint $\nu\mathcal{F}_\alpha$ coincides with behavioural
equivalence:

\begin{reptheorem}
  Let $F$ be separable by singletons, and let
  $\alpha: X\rightarrow FX$ be an $F$-coalgebra.  Then
  $\nu \mathcal{F}_{\alpha} = W_{\alpha}$, i.e.,
  $\nu \mathcal{F}_{\alpha}$ contains exactly the pairs
  $(x_0,x_1)\in X\times X$ for which the duplicator has a winning
  strategy.
\end{reptheorem}

\begin{laterproof}~
  
  \begin{description}
  \item[``$\subseteq$''] Assume that
    $(x_0,x_1)\in \nu\mathcal{F}_\alpha = R_n$. We show that $x_0\sim x_1$ and
    with \cite{km:bisim-games-logics-metric} it follows that $(x_0,x_1)\in
    W_\alpha$. We do this by constructing a coalgebra homomorphism $f$
    with $f(x_0) = f(x_1)$.

    Let $Y = E(R_n)$, the set of equivalence classes of $R_n$ and we
    define $f\colon X\to Y$, $f(x_0) = [x_0]_{R_n}$. In order to show that
    $f$ is a coalgebra homomorphism, we have to construct a coalgebra
    $\beta\colon Y\to FY$ such that $\beta\circ f = Ff\circ \alpha$.

    We define $\beta([x_0]_{R_n}) = Ff(\alpha(x_0))$ and it suffices to
    show that $\beta$ is well-defined.

    So let $(x_0,x_1)\in R_n$ and assume by contradiction that
    $t_0 = Ff(\alpha(x_0)) \neq Ff(\alpha(x_1)) = t_1$. Then, since $F$ is
    separable by singletons, we have a singleton predicate $p$
    with $Fp(t_0) \neq Fp(t_1)$. By expanding the definition we get
    $F(p\circ f)(\alpha(x_0)) \neq F(p\circ f)(\alpha(x_1))$. By
    construction $p\circ f = \chi_P$, where $P$ is an equivalence
    class of $R_n$. This is a contradiction, since
    $\mathcal{F}_\alpha(R_{n-1}) = R_n = R_{n-1}$, which indicates
    that there can not be such a $P$.
  \item[``$\supseteq$''] Whenever
    $(x_0,x_1)\notin \nu\mathcal{F}_\alpha = R_n$, we have shown in
    Proposition~\ref{prop:constr-winning-str} that the spoiler has a
    winning strategy, which implies $(x_0,x_1)\notin W_\alpha$. Hence
    $W_\alpha\subseteq \nu \mathcal{F}_\alpha$. \qed
  \end{description}
\end{laterproof}

\begin{example}
  \label{ex:classical-game-algo1} We revisit
    Example~\ref{ex:classical-game} and explain the execution of
    Algorithm~\ref{AlgorithmA_extended}. In the first iteration we
    only have to consider one predicate~$\chi_X$, and for all separated
    pairs of states $(s,t)$ we set $I(s,t)=1$ where the second
    component of $T(s,t)$ is $X$. That is, the states are
    simply divided into equivalence classes according to their
    outgoing transitions. More concretely, we obtain the separation
    of $\{1,2,3\}$ (with value $\{(a,1)\}$) from $\{4\}$ (with value
  $\{(b,1)\}$), $\{5\}$ (with value $\{(a,1),(b,1)\}$ and
  $\{6,7,8,9\}$ (with value $\emptyset$). In the second iteration the
  predicate $\chi_{\{4\}}$ is employed to separate $\{1\}$ (with value
  $\{(a,0),(a,1)\}$) from $\{2\}$ (with value $\{(a,0)\}$) and
    we get $I(1,2)=2$ with $T(1,2)=(1,\{4\})$, which also determines
    the strategy of the spoiler explained above. Similarly $\{3\}$
  can be separated from both $\{1\}$ and $\{2\}$ with the predicate
  $\chi_{\{6,7,8,9\}}$.
\end{example}
\noindent The notion of separability by singletons is needed because
the partition refinement algorithm we are using separates two states
based on a \emph{single} equivalence class of their successors,
whereas other partition refinement algorithms
(e.g.~\cite{kk:coalgebra-weighted-automata-journal}) consider
\emph{all} equivalence classes. As shown in
Example~\ref{ex:separable-singletons}, this is indeed a restriction,
however such additional assumptions seem necessary if we want to adapt
efficient bisimulation checking algorithms such as the ones by
Kanellakis/Smolka \cite{Kanellakis1990CCSEF} or Paige/Tarjan
\cite{DBLP:journals/siamcomp/PaigeT87} to the coalgebraic setting. In
fact, the Paige/Tarjan algorithm already has a coalgebraic version
\cite{DorschEA17} which operates under the assumption that the functor
is \emph{zippable}. Here we show that the related notion of
$m$-zippability is very similar to separability by singletons. (The
zippability of \cite{DorschEA17} is in fact $2$-zippability, which is
strictly weaker than $3$-zippability
\cite{w:personal,w:coalgebra-minimization}.)

\begin{definition}[zippability]
  A functor $F$ is
  \emph{$m$-zippable} if the map
  \[ F(A_1 + \dots + A_m) \xlongrightarrow{\langle F(f_1), \dots ,
      F(f_m) \rangle} F(A_1+1) \times \dots \times F(A_m+1) \] is
  injective for all sets $A_1,\dots,A_m$, where
  $f_i = \mathit{id}_{A_i} +\ ! \colon A_1+\dots+A_m \to
    A_i+1$, with $!\colon A_1+\dots+A_{i-1}+A_{i+1}+\dots+A_m\to 1$,
  is the function mapping all elements of $A_i$ to themselves and all
  other elements to $\bullet$ (assuming that $1 = \{\bullet\}$).
\end{definition}
  
\begin{replemma}
  If a functor $F$ is separable by singletons,
  then~$F$ is $m$-zippable for all~$m$. Conversely, if $F$ is
  $m$-zippable, then~$F$ is separable by singletons on all sets~$X$
  with $|X|\le m$.
\end{replemma}
  
\begin{laterproof} 
  ~
  \begin{itemize}
  \item Suppose that $F$ is separable by singletons. We need to show
    that
    \[ F(A_1 + \dots + A_m) \xlongrightarrow{\langle F(f_1), \dots ,
        F(f_m) \rangle} F(A_1+1) \times \dots \times F(A_m+1) \] is
    injective.  Hence let $t_0,t_1 \in F(A_1 + \dots + A_m)$ with
    \[\langle F(f_1), \dots , F(f_m) \rangle (t_0) = \langle
      F(f_1), \dots , F(f_m) \rangle (t_1)\] be given. The situation
    is depicted in Figure~\ref{fig:mono-func-zip} below.

    Now let $x_i\in A_i$ and consider the singleton predicate
    $\chi_{\{x_i\}} \colon A_1+\dots+A_m\to 2$, which decomposes as
    $\chi_{\{x_i\}} = h_{x_i}\circ f_i$ where $h_{x_i}\colon A_i+1\to 2$
    is the characteristic function of $x_i$ on $A_i+1$ (see
    Figure~\ref{fig:characteristic-func-zip} below).

    \begin{figure}[h]
      \centering
      \begin{minipage}[b]{.3\linewidth} 
        \centering	
        \begin{tikzpicture}
          \node (A) at (0,0) []{$F(A_1 + \dots +A_m)$}; 
          \node (AC) at (0,-2) []{$\prod \limits_{i \in  \{1, \dots, m\} }F(A_i+1) $}; 
          \node (B1) at (-2.2,-3) []{$F(A_1+1)$}; 
          \node (Bm) at (2.2,-3) []{$F(A_m+1)$};
          \node (D) at (0,-3) []{$\dots$};
          
          \draw  [->] (A) to []  node  []{$\langle F(f_1), \dots , F(f_m) \rangle $} (AC);
          \draw  [->] (A) to [bend right=40]  node  [left]{$ Ff_1 $} (B1);
          \draw  [->] (A) to [bend left=40]  node  [right]{$ Ff_m $} (Bm);
          \draw  [->] (AC) to [bend left=20]  node  [below]{$ \pi_1 $} (B1);
          \draw  [->] (AC) to [bend right=20]  node  [below]{$ \pi_m $} (Bm);
          
        \end{tikzpicture}
        \caption{}
        \label{fig:mono-func-zip}
      \end{minipage}
      \hspace{.2\linewidth}
      \begin{minipage}[b]{.4\linewidth} 
        \centering	
        \begin{tikzpicture}
          \node (A) at (0,0) []{$A_1 + \dots + A_m$}; 
          \node (AC) at (2.5,0) []{$A_i + 1$}; 
          \node (2) at (2.5,-1.5) []{$2$}; 
          
          \draw  [->] (A) to []  node  [above]{$f_i $} (AC);
          \draw  [->] (A) to []  node  [left]{$\chi_{\{x_i\}} $} (2);	
          \draw  [->] (AC) to []  node  [right]{$h_{x_i}$} (2);
          
        \end{tikzpicture}
        \caption{}
        \label{fig:characteristic-func-zip}
      \end{minipage}
    \end{figure}

    Now we can proceed as follows:
    \begin{align*}
      &\pi_i (\langle F(f_1), \dots , F(f_m) \rangle (t_0))
      = \pi_i (\langle F(f_1), \dots , F(f_m) \rangle (t_1))\\
      \Rightarrow\qquad &\ F(f_i) (t_0) =  F(f_i) (t_1)\\
      \Rightarrow\qquad &\ Fh_{x_i} (F(f_i) (t_0)) = Fh_{x_i} (F(f_i) (t_1))\\
      \Rightarrow\qquad &\ F\chi_{\{x_i\}} (t_0) = F\chi_{\{x_i\}} (t_1)
    \end{align*}
    Since this holds for all $x_i$ in $A_1+\dots+A_m$, and $F$ is
    separable by singletons, we can conclude that $t_0=t_1$.
  \item We first observe that every functor that is $m$-zippable is
    also $m'$-zippable for $m'\le m$ (just take $A_i = \emptyset$ for
    some $i$). Hence it is sufficient to prove that whenever $F$ is
    $m$-zippable, then it is separable by singletons on all sets $X$
    with $|X|=m$. So we can assume without loss of generality that
    $X = \{x_1,\dots,x_m\}$.

    We set $A_i = \{x_i\}$ and know from the premise that
    \[F(A_1 + \dots + A_m) \xlongrightarrow{\langle F(f_1), \dots,
        F(f_m) \rangle} F(A_1+1) \times \dots \times F(A_m+1) \] is
    injective (see Figure \ref{fig:mono-func-zip}).

    Let $t_0,t_1 \in FX$ and $t_0 \neq t_1$ be given. Due to the
    injectivity of the map above, we know that there exists an index
    $i$ such that $Ff_i(t_0) \neq Ff_i(t_1)$. Since $A_i+1\cong 2$,
    every $f_i$ is itself a singleton predicate and hence we witness
    the inequality of $t_0,t_1$ via a singleton. \qed
  \end{itemize}
\end{laterproof}

\subsubsection{Runtime Analysis}

We assume that $X$ is finite and that the inequalities in
  Algorithm~\ref{AlgorithmA_extended} (with respect to $\leq^F$) are
  decidable in polynomial time. Then our algorithm  terminates
  and has polynomial runtime.

In fact, if $|X|=n$, the algorithm runs through at most $n$ iterations, since
there can be at most $n$ splits of equivalence classes. In each
iteration we consider up to $n^2$ pairs of states, and in order to
decide whether a pair can be separated, we have to consider up to $n$
equivalence classes, which results in $O(n^4)$ steps (not
  counting the steps required to decide the inequalities).

For a finite label set $A$, the inequalities are decidable in
  linear time for the functors in our examples
  ($F = \mathcal{P}_f(A\times (-))$ and $F = (\mathcal{D}(-) + 1)^A)$.
We expect that we can exploit optimizations based on
\cite{Kanellakis1990CCSEF,DBLP:journals/siamcomp/PaigeT87}. In
particular one could incorporate the generalization of the
Paige-Tarjan algorithm to the coalgebraic setting \cite{DorschEA17}.

\section{Construction of Distinguishing Formulas}
\label{sec:gen_formula}

Next we illustrate how to derive a distinguishing modal formula
from the winning strategy of $S$ computed by
Algorithm~\ref{AlgorithmA_extended}. The other direction (obtaining
the winning strategy from a distinguishing formula) has been
covered in \cite{km:bisim-games-logics-metric}.

\subsection{Cone Modalities}

We focus on an on-the-fly extraction of relevant modalities, to our
knowledge a new contribution, and discuss the connection to 
other -- given -- sets of separating predicate liftings.

One way of enabling the construction of formulas is to specify the
separating set of predicate liftings $\Lambda$ in advance. But this
set might be infinite and hard to represent. Instead here we generate
the modalities while constructing the formula. We focus in particular
on what we call \emph{cone modalities}: given $v\in F2$ we take the
upward-closure of~$v$ as a modality.

We also explain how logical formulas with cone modalities can be
translated into other separating sets of modalities.

  \begin{definition}[Cone modalities]
    Let $v \in F2$. A cone modality $[\upar v]$ is given by the
    following evaluation map $\upar v: F2 \rightarrow 2$:
    \[ \upar v(u)=	\lambda(u)= \begin{cases}
        1, & \text{if }v \leq^{F} u \\
        0, & \text{otherwise}
      \end{cases}
    \]
  \end{definition}
  \noindent Under our running assumptions, these evaluation maps yield
  a separating set of predicate liftings: Since~$F$ has a separating
  set of monotone predicate liftings, it suffices to show that the
  evaluation maps are jointly injective on~$F2$. Now if $v_0\neq v_1 $
  for $v_0,v_1\in F2$, then w.l.o.g.\ $ v_0 \nleq^{F} v_1 $, since we
  require that the lifted order is anti-symmetric. Hence,
  $\upar v_0(v_0) = 1$ and $\upar v_0(v_1) = 0$.

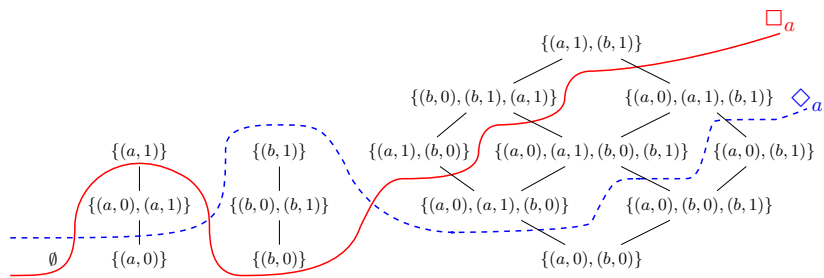
\begin{figure}[h]
  \centering	
  \resizebox{0.9\textwidth}{!}{%
    \begin{tikzpicture}
      \node (C0) at (-5,-5) [] {$\emptyset$};
      \node (C1_1) at (-3.4,-5) [] {$\{(a,0)\}$};
      \node (C1_2) at (-3.4,-4) [] {$\{(a,0),(a,1)\}$};
      \node (C1_3) at (-3.4,-3) [] {$\{(a,1)\}$};
      
      \node (C2_1) at (-0.8,-5) [] {$\{(b,0)\}$};
      \node (C2_2) at (-0.8,-4) [] {$\{(b,0),(b,1)\}$};
      \node (C2_3) at (-0.8,-3) [] {$\{(b,1)\}$};
      
      \node (C3_1) at (5,-5) [] {$\{(a,0),(b,0)\}$};
      \node (C3_2) at (3.2,-4) [] {$\{(a,0),(a,1),(b,0)\}$};
      \node (C3_3) at (7,-4) [] {$\{(a,0),(b,0),(b,1)\}$};
      \node (C3_5) at (5,-1) [] {$\{(a,1),(b,1)\}$};		
      \node (C3_4) at (5,-3) [] {$\{(a,0),(a,1),(b,0),(b,1)\}$};
      \node (C3_6) at (1.8,-3) [] {$\{(a,1),(b,0)\}$};
      \node (C3_7) at (8.2,-3) [] {$\{(a,0),(b,1)\}$};
      
      \node (C3_8) at (3,-2) [] {$\{(b,0),(b,1),(a,1)\}$};
      \node (C3_9) at (7,-2) [] {$\{(a,0),(a,1),(b,1)\}$};
      
      \node [] (Ba) at (8.5,-0.6) [red,thick, font=\fontsize{15}{15}] {$\Box_a$};
      \draw [red, thick] (-5.8,-5.3) .. controls (-5.0, -5.3) and (-4.6, -5.3) .. (-4.6, -4.4) .. controls (-4.6,-3.5) and (-3.6,-2.8) .. (-2.5,-3.5)
      .. controls (-2.1,-3.8) and (-2.1,-4.2) .. (-2.1,-4.4)
      .. controls (-2.1,-5.0) and (-1.9,-5.3) .. (-1.5,-5.3)
      .. controls (-0.9,-5.3) and (0.4,-5.3) .. (0.7,-4.4)
      .. controls (0.9,-4.0) and (1.1,-3.5) .. (1.5,-3.5)
      .. controls (1.9,-3.5) and (2.9,-3.5) .. (2.9,-3.0)
      .. controls (2.9,-3.0) and (2.9,-2.5) .. (3.3,-2.5)
      .. controls (3.3,-2.5) and (4.5,-2.5) .. (4.5,-2.0)
      .. controls (4.5,-2.0) and (4.5,-1.5) .. (5.0,-1.5)
      .. controls (5.0,-1.5) and (6.3,-1.5) .. (8.5,-0.8)
      ;
      
      \node [] (Ba) at (9.0,-2.1) [blue,thick, font=\fontsize{15}{15}] {$\Diamond_a$};
      \draw [blue, dashed,thick] 
      (-5.8,-4.6) .. controls (-2.0, -4.6) and (-1.8, -4.6) .. (-1.8, -3.0)
      .. controls (-1.8, -2.8) and (-1.8, -2.5) .. (-0.9, -2.5)	
      .. controls (-0.1, -2.5) and (0.0, -2.5) .. (0.4, -3.2)
      .. controls (0.7, -4.0) and (1.5, -4.5) .. (2.5, -4.5)
      .. controls (1.9, -4.5) and (5.0, -4.5) .. (5.0, -4.3)
      .. controls (5.4, -3.5) and (5.4, -3.5) .. (5.9, -3.5)
      .. controls (5.9, -3.5) and (5.4, -3.5) .. (6.7, -3.5)
      .. controls (6.9, -3.5) and (7.0, -3.5) .. (7.2, -2.6)
      .. controls (7.3, -2.4) and (7.0, -2.4) .. (8.4, -2.4)
      .. controls (8.4, -2.4) and (8.5, -2.4) .. (9.0, -2.2)
      ;
            
      \draw (C1_1) -- (C1_2);
      \draw (C1_2) -- (C1_3);
      
      \draw (C2_1) -- (C2_2);
      \draw (C2_2) -- (C2_3);
      
      \draw (C3_1) -- (C3_2);
      \draw (C3_1) -- (C3_3);
      
      \draw (C3_2) -- (C3_4);
      \draw (C3_2) -- (C3_6);
      
      \draw (C3_3) -- (C3_4);
      \draw (C3_3) -- (C3_7);	
      
      \draw (C3_6) -- (C3_8);
      \draw (C3_7) -- (C3_9);
      
      \draw (C3_4) -- (C3_8);
      \draw (C3_4) -- (C3_9);
      
      \draw (C3_8) -- (C3_5);
      \draw (C3_9) -- (C3_5);	
    \end{tikzpicture}	}%
  \caption{The set $F2 = \mathcal{P}_f(\{a,b\}\times 2) $ with the order $\leq^F$ (for labelled transition
    systems). The modality $\Box_a$ ($\Diamond_a$) is given by the elements above
    the solid (dashed) line.}
  \label{fig:cone_modalities_pow}
\end{figure}

\begin{example}
  We discuss modalities respectively evaluation maps in more
  detail for the functor $F = \mathcal{P}_f(A\times (-))$ (see also
  Example~\ref{ex:classical-game}). In our example, $A = \{a,b\}$. The
  set $F2$ with order $\leq^F$ is depicted as a Hasse diagram in
  Figure~\ref{fig:cone_modalities_pow}. For every element there is a
  cone modality, 16 modalities in total. It is known from the
  Hennessy-Milner theorem \cite{HennessyMilner_Logic:1985} that two
  modalities are enough: either $\Box_a,\Box_b$ (box modalities) or
  $\Diamond_a,\Diamond_b$ (diamond modalities), where for $v\in F2$,
  \begin{equation*}
    \Box_a(v)  = 
    \begin{cases}
      1 & \text{if } (a,0)\notin v \\
      0 & \text{otherwise}
    \end{cases}
    \qquad
    \Diamond_a(v)  = 
    \begin{cases}
      1 & \text{if } (a,1) \in v \\
      0 & \text{otherwise.}
    \end{cases}
  \end{equation*}
  \noindent In Figure~\ref{fig:cone_modalities_pow}, $\Box_a$
respectively $\Diamond_a$ are represented by the elements above the
two lines (solid respectively dashed).
\end{example}

\begin{figure}[h]
  \centering	
  \begin{tikzpicture}
    \small{
      \node (C0) at (-5,-3) [] {$	
        [a\mapsto \bullet,b\mapsto \bullet]
        $};
      
      \node (C1bot) at (-2,-3) [] {$
        [a\mapsto 0,b\mapsto \bullet]
        $};
      \node(C) at (-2,-2.4) [] {$\vdots$};
      
      \node (C1top) at (-2,-2) [] {$
        [a\mapsto 1,b\mapsto \bullet]
        $};

      \node (C2bot) at (1,-3) [] {$
        [a\mapsto \bullet,b\mapsto 0]
        $};
      \node(C2) at (1,-2.4) [] {$\vdots$};
      
      \node (C2top) at (1,-2) [] {$
        [a\mapsto \bullet,b\mapsto 1]
        $};
      
      \node (C3bot) at (4,-3) [] {$
        [a\mapsto 0,b\mapsto 0]
        $};
      \node(C3a) at (3.6,-2.4) [] {$\ddots$};
      \node(C3b) at (4.4,-2.4) [] {$\iddots$};
      \node(C3c) at (3.6,-2) [] {$\iddots$};
      \node(C3d) at (4.4,-2) [] {$\ddots$};
      
      \node (C3top) at (4,-1.6) [] {$
        [a\mapsto 1,b\mapsto 1]
        $};
    }
  \end{tikzpicture}
  \caption{$F2 \cong ([0,1]+1)^A$ with order $\leq^F$ (for probabilistic
    transition systems).}
  \label{fig:cone_modalities_distr}
\end{figure}
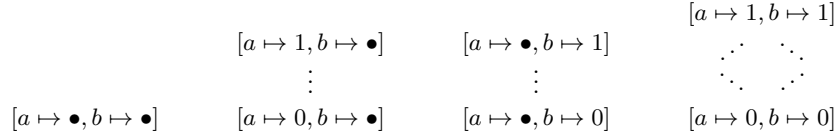

\begin{example}
  \label{ex:distr-order}
  As a second example we discuss the functor
  $ F =(\mathcal{D}(-)+1)^A$, specifying probabilistic transition
  systems. The singleton set $1 = \{\bullet\}$ denotes
  termination. Again we set $A = \{a,b\}$.

  Since $\mathcal{D}2$ is isomorphic to the interval $[0,1]$, we can
  simply represent any distribution $d\colon 2\to [0,1]$ by
  $d(1)$. Hence $F2 \cong ([0,1]+1)^A$. The partial order is
  componentwise and is depicted in
  Figure~\ref{fig:cone_modalities_distr}: it decomposes into four
  disjoint partial orders, depending on which of $a,b$ are mapped to
  $\bullet$.  The right-hand part of this partial order consists of
  function $[0,1]^A$ with the pointwise order.
  
  We will also abbreviate a map $[a\mapsto p,b\mapsto
  q]$ by $\langle a_p,b_q\rangle$.
\end{example}

\subsection{From Winning Strategies to Distinguishing Formulas}
We will now show how a winning strategy of $S$ can be transformed into
a distinguishing formula, based on cone modalities, including some
examples.

The basic idea behind the construction in Definition
\ref{def:constr_formu} is the following: Let $(x_0,x_1)$ be a pair of
states separated in the $i$-th iteration of the partition refinement
algorithm (Algorithm~\ref{AlgorithmA_extended}). This means that we
have the following situation:
$F\chi_{P}(\alpha(x_0)) \nleq^{F} F\chi_{P}(\alpha(x_1))$ (or vice
versa) for some equivalence class $P$ of $R_{i-1}$. Based on
$v = F\chi_{P}(\alpha(x_0))$ we define a cone modality
$\lambda = \ \upar v$. Now, if we can characterize $P$ by some formula
$\psi$, i.e., $\llbracket\psi\rrbracket = \chi_P$ (we will later show
that this is always possible), we can define the formula
$\phi = [\lambda]\psi$. Then it holds that:
\begin{eqnarray*}
	\llbracket\phi\rrbracket(x_0) =
	\lambda(F\llbracket\psi\rrbracket(\alpha(x_0))) = \ \upar
	v(F\chi_P(\alpha(x_0))) = 1 \\
	\llbracket\phi\rrbracket(x_1) =
	\lambda(F\llbracket\psi\rrbracket(\alpha(x_1))) = \ \upar
	v(F\chi_P(\alpha(x_1))) = 0
\end{eqnarray*}
That is we have $x_0\models \phi$ and $x_1\not\models \phi$, which means
that we have constructed a distinguishing formula for $x_0,x_1$.

\smallskip

First, we describe how a winning strategy for the spoiler for a pair
$(x_0,x_1)$ is converted into a formula and then prove that this formula
distinguishes $x_0,x_1$.

\begin{definition}
	\label{def:constr_formu}
	Let $x_0\nsim x_1$ (equivalently $(x_0,x_1)\notin R_n$), and
        let $(T,I)$ be the winning strategy for the spoiler computed
        by Algorithm~\ref{AlgorithmA_extended}. We construct a formula
        $\varphi_{x_0,x_1}$ as follows: assume that
        $T(x_0,x_1) = (s,P)$ where $s=x_0$. Then set
        $v=F\chi_{P}(\alpha(x_0))$, $\lambda= \ \upar v$ and
        $\varphi_{x_0,x_1}= [\lambda]\phi$, where $\phi$ is
        constructed by recursion as follows:
	\begin{itemize}
        \item $I(x_0,x_1)=1$: \qquad $\phi = \mathit{tt}$
        \item $I(x_0,x_1)>1$: \qquad
          $\phi= \bigvee_{x_0'\in P} \big(
          \bigwedge_{x_1' \in \ X\setminus P}\;
          \varphi_{x_0',x_1'}\big)$
	\end{itemize}
	If $s = x_1$, then we set $v = F\chi_P(\alpha(x_1))$ and
        $\phi_{x_0,x_1} = \lnot [\lambda]\phi$ instead.  The recursion
        terminates because $I(x_0',x_1') < I(x_0,x_1)$ (since $P$ is
        an equivalence class of $R_{i-1}$ where $i=I(x_0,x_1)$).
      \end{definition}

\begin{repproposition}
	\label{prop:constr-disting-for} Let $\alpha: X \to FX$ be a
	coalgebra and assume that we have computed $R_n,T,I$ with Algorithm
	\ref{AlgorithmA_extended}. Then, given $(x_0,x_1) \notin R_n$, the
	construction in Definition \ref{def:constr_formu} yields a formula
	$\varphi_{x_0,x_1} \in \mathcal{L}^{\kappa}(\Lambda)$ such that
	$x_0\vDash \varphi_{x_0,x_1}$ and $x_1 \nvDash \varphi_{x_0,x_1}$.
\end{repproposition}

\begin{laterproof}
	We prove this by induction over $i=I(x_0,x_1)$:
	\begin{description}
		\item[$i=1:$] $x_0,x_1$ have been separated at Step~$1$, since
		$F\chi_X (\alpha(x_0)) \nleq^F F\chi_X(\alpha(x_1))$, where
		$T(x_0,x_1) = (x_0,X)$ (or vice versa), because $X$ is the only
		equivalence class so far. Note also that $\llbracket
		\mathit{tt}\rrbracket = X$.
		
		We set $v = F\chi_X (\alpha(x_0))$, $\lambda = \upar v$ and we
		have
		\begin{eqnarray*}
			\llbracket \phi_{x_0,x_1}\rrbracket(x_0) & = & \lambda(F\llbracket
			\phi\rrbracket(\alpha(x_0))) =
			\lambda(F\chi_X(\alpha(x_0))) = \lambda(v) = 1 \\
			\llbracket \phi_{x_0,x_1}\rrbracket(x_1) & = & \lambda(F\llbracket
			\phi\rrbracket(\alpha(x_1))) =
			\lambda(F\chi_X(\alpha(x_1))) = 0
		\end{eqnarray*}
		Hence $x_0\models \phi_{x_0,x_1}$ and $x_1\not\models\phi_{x_0,x_1}$.
		
		In the case where $T(x_0,x_1) = (x_1,X)$, we have $v = F\chi_X
		(\alpha(x_1))$, $\lambda = \upar v$ and we obtain
		\begin{eqnarray*}
			\llbracket [\lambda]\phi\rrbracket(x_0) & = & \lambda(F\llbracket
			\phi\rrbracket(\alpha(x_0))) =
			\lambda(F\chi_X(\alpha(x_0))) = 0 \\
			\llbracket [\lambda]\phi\rrbracket(x_1) & = & \lambda(F\llbracket
			\phi\rrbracket(\alpha(x_1))) =
			\lambda(F\chi_X(\alpha(x_1))) = \lambda(v) = 1
		\end{eqnarray*}
		Hence again $x_0\models \phi_{x_0,x_1}$ and $x_1\not\models\phi_{x_0,x_1}$.
			
		\item[$i\to i+1:$] Due to the induction hypothesis we can assume
		that the $\phi_{x_0',x_1'}$ are distinguishing formulas for $(x_0',x_1')$
		with $I(x_0',x_1') < i+1$.
		
		First, we show that $\llbracket\phi\rrbracket = P$. 
		\begin{itemize}
			\item Let $z\in P$. Then there exists an $x_0'\in P$ (namely $x_0'=z$)
			such that $z\models \phi_{x_0',x_1'}$ for all $x_1'\notin
			P$. Furthermore, by construction of $\phi_{x_0',x_1'}$ it holds that
			$x_1'\not\models \phi_{x_0',x_1'}$. This means that
			$z\models \bigwedge_{x_1'\in X\backslash P} \phi_{x_0',x_1'}$ and also
			$z\models\bigvee_{x_0'\in P} \bigwedge_{x_1'\in X\backslash P}
			\phi_{x_0',x_1'} = \phi$.
			\item Let $z\notin P$. Then for every $x_0'\in P$ there exists an
			$x_1'\notin P$ (namely $x_1' = z$) such that
			$z\not\models \phi_{x_0',x_1'}$. Hence
			$z\not\models \bigwedge_{x_1'\in X\backslash P}
			\phi_{x_0',x_1'}$. Since this is true for every such $x_0'$ we also
			have
			$z\not\models\bigvee_{x_0'\in P} \bigwedge_{x_1'\in X\backslash P}
			\phi_{x_0',x_1'} = \phi$.
		\end{itemize}
		Assume that $T(x_0,x_1) = (x_0,P)$ (the case $T(x_0,x_1) = (x_1,P)$ can be
		handled analogously as for $i=1$). Hence we know that
		$F\chi_P (\alpha(x_0)) \nleq^F F\chi_P(\alpha(x_1))$. 
		
		We set $v = F\chi_P (\alpha(x_0))$, $\lambda = \upar v$ and we
		have
		\begin{eqnarray*}
			\llbracket \phi_{x_0,x_1}\rrbracket(x_0) & = & \lambda(F\llbracket
			\phi\rrbracket(\alpha(x_0))) =
			\lambda(F\chi_P(\alpha(x_0))) = \lambda(v) = 1 \\
			\llbracket \phi_{x_0,x_1}\rrbracket(x_1) & = & \lambda(F\llbracket
			\phi\rrbracket(\alpha(x_1))) =
			\lambda(F\chi_P(\alpha(x_1))) = 0
		\end{eqnarray*}
		Hence $x_0\models \phi_{x_0,x_1}$ and $x_1\not\models\phi_{x_0,x_1}$.
	
		\qed
	\end{description}	
\end{laterproof}
\noindent We next present an optimization of the construction in
Definition~\ref{def:constr_formu}, inspired by
\cite{c:automatically-explaining-bisim}.  In the case $I(x_0,x_1) > 1$
one can pick an arbitrary $x_0'\in P$ and keep only one element of the
disjunction.

In order to show that this simplification is permissible, we need the
following lemma.

\begin{replemma}
	\label{lem:dist-formula-upto-step-i}
	Given two states $(x_0,x_1) \notin R_n$ and a distinguishing formula
	$\varphi_{x_0,x_1}$ based on Definition \ref{def:constr_formu}. Let
	$(x_0',x_1')$ be given such that $I(x_0',x_1') > I (x_0,x_1)$. Then
	$x_0' \vDash \varphi_{x_0,x_1}$ if and only if $x_1' \vDash \varphi_{x_0,x_1} $.
\end{replemma}

\begin{laterproof}
	We have to distinguish two different cases for $I(x_0',y_1')$
	\begin{description}
		\item[$I(x_0',x_1')=1$:] this can not be true since we require
		$I(x_0',x_1') > I(x_0,x_1) \ge 1$.
		\item[$I(x_0',x_1')>1$:] For any $(x_0,x_1)$ with $I (x_0,x_1) < I(x_0',x_1')$ we
		have $x_0 \vDash \varphi_{x_0,x_1}$ and $x_1 \nvDash \varphi_{x_0,x_1} $ where
		$\varphi_{x_0,x_1} = [\lambda]\phi$,
		$\lambda =\ \upar F\chi_P (\alpha(x_0))$ and $T(x_0,x_1)= (x_0,P)$ (the
		case $T(x_0,x_1) = (x_1,P)$ is analogous). Furthermore, the semantics of
		$\phi$ is $\llbracket\phi \rrbracket = \chi_P $ (for details we
		refer to the proof of Proposition
		\ref{prop:constr-disting-for}). Now, assume without loss of
		generality that the following holds
		\begin{align*}
		1=&\ \llbracket \varphi_{x_0,x_1} \rrbracket (x_0')= \lambda ( F\chi_P
		(\alpha(x_0'))) \\
		0=&\ \llbracket \varphi_{x_0,x_1} \rrbracket (x_1')= \lambda ( F\chi_P
		(\alpha(x_1')))
		\end{align*}
		Due to Proposition~\ref{prop:separating-pl-antisym-ji} $\lambda$
		is monotone. Therefore, the above assumption implies
		$F\chi_P (\alpha(x_0')) \nleq^{F} F\chi_P (\alpha(x_1')) $. But this
		yields a contradiction, since then $x_0',x_1'$ would have been
		separated in a Step~$i \le I(x_0,x_1) < I(x_0',x_1')$. \qed
	\end{description}
\end{laterproof}

\noindent Now we can show that we can replace the formula $\phi$ from Definition
\ref{def:constr_formu} by a simpler formula $\phi'$.

\begin{replemma}
	Let $(x_0,x_1)\notin R_i$ and let $P$ be an equivalence class of
	$R_{i-1}$. Furthermore let
	\[
	\phi' =  \bigwedge_{\substack{x_1' \in \ X\setminus P}}
	\varphi_{x_0',x_1'}
	\]
	for some $x_0'\in P$. Then $\llbracket \phi' \rrbracket = \chi_{P}$.
\end{replemma}

\begin{laterproof}
	Clearly $\llbracket \phi'\rrbracket \le \llbracket
	\phi\rrbracket = \chi_P$.
	
	We now have to show that the other inequality holds as well, so let
	$z\in P$. Furthermore let $x_1'$ be arbitrary such that $x_1'\notin P$.
	Since $z,x_0'\in P$ and $x_1'\notin P$, where $P$ is an equivalence
	class, we know that $I(z,x_0') > I(x_0',x_1')$ (possibly even
	$I(z,x_0') = \infty$). Hence, by
	Lemma~\ref{lem:dist-formula-upto-step-i} we have that
	$z\models \phi_{x_0',x_1'}$ if and only if $x_0'\models \phi_{x_0',x_1'}$. And
	since the latter holds, we have $z\models \phi_{x_0',x_1'}$.
	
	Hence
	$z\models \bigwedge_{x_1'\in X\backslash P} \phi_{x_0',x_1'} = \phi$. In
	summary, we get $\chi_P\le \llbracket \phi\rrbracket$.
	\qed
\end{laterproof}

\noindent Finally, we can simplify our construction described in Definition
\ref{def:constr_formu} to only one inner conjunction.

\begin{corollary}
	\label{def:simpl_constr_formu}
	We use the construction of $\phi_{x_0,x_1}$ as described in
	Definition~\ref{def:constr_formu} with the only modification that
	for $I(x_0,x_1) > 1$ the formula $\phi$ is replaced by

	\[
	\phi' =  \bigwedge_{\substack{x_1' \in \ X\setminus P}}
	\varphi_{x_0',x_1'}
	\]
	for some $x_0'\in P$.
	Then this yields a formula $\varphi_{x_0,x_1}$ such
	that $x_0\vDash \varphi_{x_0,x_1}$ and $x_1 \nvDash \varphi_{x_0,x_1}$.
\end{corollary}

\noindent A further optimization takes only one representative $x_1'$ from every
equivalence class different from $P$.

\smallskip

We now explore two slightly more complex examples.

\begin{example}
  Take the coalgebra for the functor $F = (\mathcal{D}(-) + 1)^A$
  depicted in Figure~\ref{fig:exa-prob-ts}, with $A = \{a,b\}$ and set
  $X = \{1,\dots,5\}$ of states. For instance,
  $\alpha(3) = [a\mapsto \delta_3, b\mapsto \bullet]$ where $\delta_3$
  is the Dirac distribution. This is visualized by drawing an arrow
  labelled $a,1$ from $3$ to $3$ and omitting $b$-labelled arrows.
  
  We explain only selected steps of the construction: In the first
  step, the partition refinement algorithm
  (Algorithm~\ref{AlgorithmA_extended}) separates $1$ from $3$ (among
  other separations), where the spoiler strategy is given by
  $T(1,3) = (1,X)$. In order to obtain a distinguishing formula, we
  determine $v = F\chi_X(\alpha(1)) = \langle a_1,b_1\rangle$ (using
  the abbreviations explained in Example~\ref{ex:distr-order}) and
  obtain $\phi_{1,3} = [\upar\langle a_1,b_1\rangle]\mathit{tt}$. In
  fact, this formula also distinguishes $1$ from $4$, hence
  $\phi_{1,3} = \phi_{1,4}$.
  If, on the other hand, we want to distinguish $3,4$, we 
  obtain
  $\phi_{3,4} = [\upar\langle a_1,b_\bullet\rangle]\mathit{tt}$.

  After the first iteration, we obtain the partition
  $\{1,2,5\}, \{3\}, \{4\}$. Now we consider states $1,2$ which can be
  separated by playing $T(2,1) = (2,\{1,2,5\})$, since $5$ behaves
  differently from $3$. Again we compute
  $v = F\chi_P(\alpha(2)) = \langle a_{1},b_{0.8}\rangle $ (for
  $P = \{1,2,5\}$) and obtain
  $\phi_{2,1} = [\upar\langle a_{1},b_{0.8}\rangle](\phi_{1,3}\land
  \phi_{1,4})$. 
  Here we picked $1$ as the representative of its equivalence class.

  In summary we obtain $\phi_{2,1}=[\upar\langle a_{1},b_{0.8}\rangle][\upar\langle
  a_1,b_1\rangle]\mathit{tt}$, which is satisfied by~$2$ but not by~$1$.
\end{example}

\begin{figure}[h]
	\centering
	\begin{minipage}{.4\linewidth} 
          \hspace{-0.5cm}
                \scalebox{0.9}{
		\begin{tikzpicture}
		
		\node (S1) at (0,-2) [circle,draw]{$1$}; 			
		\node (S3) at (-1.2,-3.2) [circle,draw]{$3$};	  
		\node (S5) at (1.2,-3.2) [circle,draw]{$4$}; 
		
		\draw  [->] (S1) to [bend left=20]  node  [right]{$ $} (S5);	
		\draw  [->] (S1) to [bend right=20] node  [left]{$b,0.8$} (S3);	
		\draw  [->] (S1) to [bend left=20] node  [right]{$a,0.7$} (S3);		
		\path
		(S3) edge [loop below] node {$a,1$} (S3);
		
		\path	
		(S1) edge [loop above] node {$a,0.3$} (S1);	
		 		
		\node (T1) at (2.4,-2) [circle,draw]{$2$}; 			
		\node (T6) at (3.6,-3.2) [circle,draw]{$5$};  
		
		\draw  [->] (T1) to [bend right=20]  node  [left]{$ $} (S5);	
		\draw  [->] (T1) to [bend left=20] node  [right]{$a,0.7$} (T6);
		\draw  [->] (T1) to [bend right=20] node  [below]{$ $} (T6);
		\path
		(T6) edge [loop below] node {$a,1$} (T6);
		\path
		(T6) edge [loop right] node {$b,1$} (T6);
		\path
		(T1) edge [loop above] node {$a,0.3$} (T1);	
				
		\node (s1) at (0.9,-2.2) []{$b,0.2$}; 
		\node (t1) at (2.0,-2.6) []{$b,0.2$}; 
		\node (t2) at (2.6,-3.0) []{$b,0.8$};
		\end{tikzpicture}}		
		\caption{Probabilistic transition system}	
		\label{fig:exa-prob-ts}	
	\end{minipage}
	\hspace{.07\linewidth}
	\begin{minipage}{.4\linewidth} 
    \scalebox{0.8}{
      \hspace{-1cm}
    \begin{tikzpicture}
    
    \node (S1) at (-1.0,0) [circle,draw]{$1$}; 			
    \node (S2) at (1.0,0) [circle,draw]{$2$};
    \node (S3) at (0.0,-1.4) [circle,draw]{$3$};  
    \node (S4) at (-2.0,-1.4) [circle,draw]{$4$}; 
    \node (S5) at (2.0,-1.4) [circle,draw]{$5$}; 
    \node (S6) at (-2.8,-2.8) [circle,draw]{$6$};
    \node (S7) at (-1.0,-2.8) [circle,draw]{$7$};
    \node (S8) at (1.0,-2.8) [circle,draw]{$8$};
    \node (S9) at (3.0,-2.8) [circle,draw]{$9$};
    
    \draw  [->] (S1) to [bend left=20]  node  [right]{$a$} (S3);
    \draw  [->] (S2) to [bend right=20]  node  [below]{$a$} (S4);
    \draw  [->] (S2) to [bend left=20]  node  [right]{$a$} (S5);
    
    \draw  [->] (S3) to [bend left=20]  node  [right]{$b$} (S8);
    \draw  [->] (S3) to [bend right=20]  node  [left]{$b$} (S7);
    \draw  [->] (S4) to [bend right=20]  node  [left]{$b$} (S6);
    \draw  [->] (S4) to [bend left=20]  node  [above]{$b$} (S7);
    \draw  [->] (S4) to [bend left=20]  node  [pos=0.3,above]{$b$} (S8);
    \draw  [->] (S5) to [bend right=20]  node  [above]{$b$} (S8); 
    \draw  [->] (S5) to [ bend right=20]  node [pos=0.3,above] {$b$}  (S7); 	
    \draw  [->] (S5) to [bend left=20]  node  [right]{$b$} (S9); 
    \path
    (S6) edge [loop left] node {$e$} (S6);
    \path	
    (S7) edge [loop left] node {$c$} (S7);	
    \path	
    (S8) edge [loop right] node {$d$} (S8);	
    \path	
    (S9) edge [loop right] node {$f$} (S9);	  	
    \end{tikzpicture}	}	
    \caption{Non-deterministic transition system.}	
    \label{fig:exa-nfa-conj-neg}
	\end{minipage}
\end{figure}

\begin{example}
  We will now give an example where conjunction is required to obtain
  the distinguishing formula. We work with the coalgebra for the functor
  $F = \mathcal{P}_f(A\times (-))$ depicted in
  Figure~\ref{fig:exa-nfa-conj-neg}, with $A = \{a,b,c,d,e,f\}$ and
  set $X = \{1,\dots,9\}$ of states.
  
  We explain only selected steps: In the first step, the partition
  refinement separates $6$ from $7$ (among other separations), where
  the spoiler strategy is given by $T(6,7) = (6,X)$. As explained
  above, we determine $v = F\chi_X(\alpha(6)) = \{(e,1)\}$ and obtain
  $\phi_{6,7} = [\upar\{(e,1)\}]\mathit{tt}$. In fact, this formula
  also distinguishes $6$ from all other states, so we denote it by
  $\phi_{6,*}$.

  Next, we consider the states $3,4$, where the possible moves of $3$ are
  a proper subset of the moves of $4$. Hence the spoiler strategy is
  $T(3,4) = (4,\{6\})$, i.e., the spoiler has to move to state $6$,
  which is not reachable from $3$. Again we compute
  $v = F\chi_P(\alpha(4)) = \{(b,1),(b,0)\}$ (for $P = \{6\}$) and
  obtain $\phi_{3,4} = \linebreak \lnot
  [\upar\{(b,1),(b,0)\}]\phi_{6,*}$. Note that this time we have to
  use negation, since the spoiler moves from the second state in the
  pair.

  Finally, we consider the states $1,2$, where the spoiler strategy is
  $T(1,2) = (1,\{3\})$. We compute
  $v = F\chi_P(\alpha(1)) = \{(a,1)\}$ (for $P = \{3\}$) and obtain
  $\phi_{1,2} = [\upar\{(a,1)\}]\big(\bigwedge_{x\in\{1,2,4,\dots,9\}}
  \phi_{3,x}\big)$. In fact, here it is sufficient to consider $x = 4$
  and $x = 5$, resulting in the following distinguishing formula:
  \[ [\upar\{(a,1)\}]  \big(\neg
    [\upar\{(b,0),(b,1)\}][\upar\{(e,1)\}]\mathit{tt} \ \wedge \ \neg
    [\upar\{(b,0),(b,1)\}][\upar\{(f,1)\}]\mathit{tt}\big). \]
\end{example}

\subsection{Recoding Modalities}
\label{sec:recode}

Finally, we will show under which conditions one can encode cone
modalities into given generic modalities, determined by a separating set of
predicate liftings $\Lambda$, not necessarily monotone. We first need
the notion of strong separation.

\begin{definition}
  Let $\Lambda$ be a separating set of predicate liftings of the form
  $\lambda\colon F2\to 2$. We call $\Lambda$ \emph{strongly
    separating} if for every $t_0 \neq t_1$ with $t_0,t_1\in F2$ there exists
  $\lambda\in\Lambda$ such that $\lambda(t_0) \neq \lambda(t_1)$.
\end{definition}

\noindent We can generate a set of strongly separating predicate liftings from
every separating set of predicate liftings. 

\begin{replemma}
  Let $\Lambda$ be a separating set of predicate liftings. Furthermore we denote the four functions on $2$ by $\mathit{id}_2$,
  $\mathit{one}$ (constant $1$-function), $\mathit{zero}$ (constant
  $0$-function) and $\mathit{neg}$ ($\mathit{neg}(0) = 1$,
  $\mathit{neg}(1) = 0$).

  Then
  \[ \Lambda' = \{ \lambda, \lambda\circ F\mathit{one}, \lambda\circ
    F\mathit{zero}, \lambda\circ F\mathit{neg} \mid
    \lambda\in\Lambda\} \]
  is a set of strongly separating predicate liftings.

  Furthermore for every formula $\phi$ we have that
  \begin{eqnarray*}
    && [\lambda\circ F\mathit{one}]\phi \equiv [\lambda]\mathit{tt}
    \qquad
    {}[\lambda\circ F\mathit{zero}]\phi  \equiv  [\lambda]\mathit{ff}
    \qquad 
    {}[\lambda\circ F\mathit{neg}]\phi  \equiv  [\lambda](\lnot \phi)
  \end{eqnarray*}
\end{replemma}

\begin{laterproof}
  Let $t_0,t_1\in F2$ with $t_0\neq t_1$. According to the definition
  of separation there must be a predicate $p\colon 2\to 2$ such
  that $\lambda(Fp(t_0)) \neq \lambda(Fp(t_1))$. Since there are only
  four such functions, $p$ must be one of $\mathit{id}_2$,
  $\mathit{one}$, $\mathit{zero}$, $\mathit{neg}$ and we immediately
  obtain that $\Lambda'$ is strongly separating.

  In addition we have that, given a coalgebra $\alpha\colon X\to FX$:
  \begin{eqnarray*}
    \llbracket [\lambda\circ F\mathit{one}]\phi\rrbracket & = &
    \lambda \circ F\mathit{one}\circ F\llbracket \phi\rrbracket \circ
    \alpha = \lambda \circ F(\mathit{one}\circ \llbracket
    \phi\rrbracket) \circ \alpha \\
    & = & \lambda \circ F\llbracket
    \mathit{tt}\rrbracket\circ \alpha = \llbracket
    [\lambda]\mathit{tt} \rrbracket \\
   \llbracket [\lambda\circ F\mathit{zero}]\phi\rrbracket & = &
    \lambda \circ F\mathit{zero}\circ F\llbracket \phi\rrbracket \circ
    \alpha = \lambda \circ F(\mathit{zero}\circ \llbracket
    \phi\rrbracket) \circ \alpha \\
    & = & \lambda \circ F\llbracket
    \mathit{ff}\rrbracket\circ \alpha = \llbracket
    [\lambda]\mathit{ff} \rrbracket \\
   \llbracket [\lambda\circ F\mathit{neg}]\phi\rrbracket & = &
    \lambda \circ F\mathit{neg}\circ F\llbracket \phi\rrbracket \circ
    \alpha = \lambda \circ F(\mathit{neg}\circ \llbracket
    \phi\rrbracket) \circ \alpha \\
    & = & \lambda \circ F\llbracket
    \lnot \phi\rrbracket\circ \alpha = \llbracket
    [\lambda]\lnot\phi \rrbracket 
  \end{eqnarray*}
  \qed
\end{laterproof}

\noindent This means that we can still express the new modalities with the
previous ones. $\Lambda'$ is just an auxiliary construct that helps us
to state the following proposition. The construction of $\Lambda'$
from $\Lambda$ was already considered in
\cite[Definition~24]{SCHRODER2008230}, where it is called
\emph{closure}.

\begin{repproposition}
  \label{prop:general-trans-of-form}
  Suppose that $F2$ is finite, and let $\Lambda$ be a strongly
  separating set of predicate liftings. Moreover, let $v \in F2$, and
  let $\varphi$ be a formula.  For $u\in F2$, we write
  $\Lambda_u = \{\lambda \in \Lambda\mid \lambda(u) = 1\}$.
  Then
  \[
    [\upar v] \varphi \equiv \bigvee_{v\le^F u} \big(
    \bigwedge_{\lambda\in \Lambda_u} [\lambda]\phi \land
    \bigwedge_{\lambda\notin \Lambda_u} \lnot[\lambda]\phi \big).
  \]
\end{repproposition}

\begin{laterproof}
  First observe that since $\Lambda$ is strongly separating, every
  $u\in F2$ is characterized uniquely by $\Lambda_u$.
  
  Let $\alpha\colon X\to FX$ be a coalgebra. We set
  $\psi_u = \bigwedge_{\lambda\in \Lambda_u} [\lambda]\phi \land
  \bigwedge_{\lambda\notin \Lambda_u} \lnot[\lambda]\phi$ and we first
  show that
  \[ x \models \psi_u \iff u = F\llbracket\phi\rrbracket(\alpha(x))
  \]
  \begin{itemize}
  \item[$\Rightarrow$:] Assume that $x\models \psi_u$. This means that
    for every $\lambda\in \Lambda_u$ we have that
    $\lambda(F\llbracket\phi\rrbracket(\alpha(x))) = 1$ and for every
    $\lambda\notin \Lambda_u$ we have that
    $\lambda(F\llbracket\phi\rrbracket(\alpha(x))) = 0$. This means
    that $u$ and $F\llbracket\phi\rrbracket(\alpha(x))$ are both
    characterized by $\Lambda_u$ and from the strong separation
    property it follows that they are equal, i.e.,
    $u = F\llbracket\phi\rrbracket(\alpha(x))$.
  \item[$\Leftarrow$:] Assume that
    $u = F\llbracket\phi\rrbracket(\alpha(x))$. Then for every
    $\lambda\in\Lambda_u$ we have that
    $\llbracket[\lambda]\phi\rrbracket(x) =
    \lambda(F\llbracket\phi\rrbracket(\alpha(x))) = \lambda(u) =
    1$. For every $\lambda\notin\Lambda_u$ we obtain
    $\llbracket[\lambda]\phi\rrbracket(x) = 0$. Everything combined,
    we have $\llbracket \psi_u\rrbracket(x) = 1$ and hence
    $x\models \psi_u$.
  \end{itemize}
  We can conclude the proof by observing that
  \begin{eqnarray*}
    && x\models [\upar v]\phi \iff v\le^F F\llbracket
    \phi\rrbracket(\alpha(x)) \iff \exists u\colon \left(v\le^F
      u\land u = F\llbracket \phi\rrbracket(\alpha(x))\right) \\
    && \iff 
    \exists u\colon \left(v\le^F u\land x\models \psi_u \right) \iff
    x\models \bigvee_{v\le^F u} \psi_u 
  \end{eqnarray*}
  \qed
\end{laterproof}
\noindent By performing this encoding inductively, we can transform a
formula with cone modalities into a formula with modalities
in~$\Lambda$. The encoding preserves negation and conjunction, only
the modalities are transformed.

\begin{example}
  We come back to labelled transition systems and the functor
  $F = \mathcal{P}_f(A\times (-))$, with $A = \{a,b\}$. In this case the set $\{\Box_a,\Box_b,\Diamond_a,\Diamond_b\}$ of
  predicate liftings is strongly separating. 

  Now let $v = \{(a,0),(b,1)\} \in \mathcal{P}_f(A\times 2)$. We show
  how to encode the corresponding cone modality using only box and
  diamond:
  \begin{eqnarray*}
    [\upar v] \varphi & \equiv & (\lnot \Box_a\phi \land \Box_b
    \phi
    \land \lnot\Diamond_a \phi \land \Diamond_b \phi) 
    \lor (\lnot \Box_a\phi \land \Box_b \phi \land \Diamond_a
    \phi \land \Diamond_b \phi) \\
    && \mbox{} \lor (\Box_a\phi \land \Box_b \phi \land \Diamond_a
    \phi \land \Diamond_b \phi)    
  \end{eqnarray*}
  The first term describes $\{(a,0),(b,1)\}$, the second
  $\{(a,0),(a,1),(b,1)\}$ and the third $\{(a,1),(b,1)\}$.
\end{example}
\noindent Note that we cannot directly generalize
Proposition~\ref{prop:general-trans-of-form} to the case where $F2$ is
infinite. The reason for this is that the disjunction over all
$u\in F2$ such that $v \leq^{F} u$ might violate the cardinality
constraints of the logic. Hence we will consider an alternative, where
the re-coding works only under certain assumptions. We will start with
the following example.

\begin{example}
  \label{ex:counter-prop-conver-formu}	
  Consider the functor $ F =(\mathcal{D}(-)+1)^A$ (see also
  Example~\ref{ex:distr-order}) and the corresponding (countable)
  separating set of (monotone) predicate liftings
  \[ \Lambda = \{ \lambda_{(a,q)}\colon F2\to 2 \mid a \in A, q\in
    [0,1]\cap \mathbb{Q}\} \cup \{\lambda_{(a,\bullet)}\mid a\in
    A\} \] where $\lambda_{(a,q)}(v)=1$ if $v(a)\in\mathbb{R}$ and
  $v(a)\geq q$ and $\lambda_{(a,\bullet)} = 1$ if $v(a) =
  \bullet$. Here, $[\lambda_{(a,q)}]\phi$ indicates that we do not
  terminate with~$a$, and the probability of reaching a state
  satisfying~$\phi$ under an $a$-transition is at least~$q$, and a
  modality $[\lambda_{(a,\bullet)}]$ ignores its argument formula, and
  tells us that we terminate with~$a$.
\end{example}

\noindent 
The disjunction $ \bigvee_{v\le^F u} $ in the construction of
$[\upar v] \varphi$ in Proposition~\ref{prop:general-trans-of-form} is
in general uncountable and may hence fail to satisfy the cardinality
constraints of the logic.  However, we can exploit certain properties
of this set of predicate liftings, in order to re-code modalities.

\begin{replemma}
  \label{lem:properties-distr-cone}
  Let $F$ be the functor with $ F =(\mathcal{D}(-)+1)^A$ and let
  $\Lambda$ be the separating set of predicate liftings from
  Example~\ref{ex:counter-prop-conver-formu}. Then
  \begin{equation}\label{eq:mod-approx}
    \upar v = \bigcap_{\lambda\in\Lambda,\lambda(v)=1 } \lambda\quad\text{for all $v\in F2$.}
  \end{equation}
\end{replemma}

\begin{laterproof}~

  \begin{description}
  \item[``$\subseteq$''] Let $u\in F2$ with $u\in \upar v$, i.e.,
    $v\le^F u$. Whenever $\lambda(v) = 1$ we also have
    $\lambda(u) = 1$ due to the monotonicity of the predicate liftings
    (cf. Proposition~\ref{prop:ev-monotone}) and hence
    $u\in\hat{\lambda}$. Since this holds for all such $\lambda$, we
    can conclude that
    $u\in \bigcap_{\lambda\in\Lambda,\lambda(v)=1 }\hat{\lambda}$.
  
  \item[``$\supseteq$''] Now suppose by contradiction that we have
    $u \in F2$ with $v \nleq^{F} u$ and
    $u \in \bigcap\limits_{\lambda\in\Lambda,
      \lambda(v)=1}\hat{\lambda}$.

    There are three cases which may cause $v \nleq^{F} u$, in
    particular they are distinguished by a specific $a\in A$:
    \begin{itemize}
    \item $v(a),u(a)\in \mathbb{R}$, but $v(a) \nleq u(a)$, which
      implies $u(a) < v(a)$. However, there exists
      $q\in [0,1]\cap \mathbb{Q}$ with $u(a) < q \le v(a)$ and for the
      corresponding modality $\lambda_{(a,q)}\in\Lambda$ we have
      $\lambda_{(a,q)}(u) = 0$, $\lambda_{(a,q)}(v) = 1$ and hence
      $u\not\in \bigcap\limits_{\lambda\in\Lambda,
        \lambda(v)=1}\hat{\lambda}$.
    \item $v(a) \in \mathbb{R}$, $u(a)=\bullet$: Now take any
      $q\in [0,1]\cap \mathbb{Q}$ with $q\le v(a)$. We use the
      modality $\lambda_{(a,q)}$, for which we have
      $\lambda_{(a,q)}(u) = 0$, $\lambda_{(a,q)}(v) = 1$ and the proof
      proceeds as before.
    \item $v(a)=\bullet$, $u(a)\in\mathbb{R}$: Now we take the
      modality $\lambda_{(a,\bullet)}$, for which we have
      $\lambda_{(a,\bullet)}(u) = 0$, $\lambda_{(a,\bullet)}(v) = 1$
      and again the proof proceeds as before.
    \end{itemize}
    \qed
  \end{description}
\end{laterproof}

\noindent Note that this property does not hold for the $\Box$ and $\Diamond$
modalities for the functor $F = \mathcal{P}_f(A\times (-))$. This can be
seen via Figure~\ref{fig:cone_modalities_pow}, where the upward
closure of $\{(b,0)\}$ contains three elements. However, $\{(b,0)\}$
is only contained in the modality $\Box_a$ (and no other modality),
which does not coincide with the upward-closure of $\{(b,0)\}$.
        
The following proposition, which relates to the well-known fact that
predicate liftings are closed under infinitary Boolean combinations
(e.g.~\cite{SCHRODER2008230}), provides a recipe for transforming cone
modalities $\upar v$ into given modalities~$\Lambda$
satisfying~\eqref{eq:mod-approx} as in
Lemma~\ref{lem:properties-distr-cone}:

\begin{repproposition}
  Given a set $\Lambda'\subseteq \Lambda$ of predicate liftings,
  understood as subsets of~$F2$, we have
  \[ [\bigcap_{\lambda\in\Lambda'} \lambda]\phi \equiv
    \bigwedge_{\lambda\in\Lambda'}[\lambda]\phi. \]
\end{repproposition}

\begin{laterproof}~
  \begin{description}
  \item[``$\subseteq$''] Let
    $x\vDash [\bigcap_{\lambda\in\Lambda'} \lambda]\varphi $, which
    implies that
    $(\bigcap_{\lambda\in\Lambda'} \lambda)(F \llbracket \varphi
    \rrbracket (\alpha(x))) = 1$. From this we conclude that
    $\lambda(F \llbracket \varphi \rrbracket (\alpha(x))) = 1$ for all
    $\lambda\in\Lambda'$, $x\models [\lambda]\phi$. And finally we
    have $x\models \bigwedge_{\lambda\in\Lambda'} [\lambda]\phi$.
  \item[``$\supseteq$''] Let
    $x \vDash \bigwedge_{\lambda\in\Lambda'} [\lambda]\phi$, which
    means that $x\vDash [\lambda]\phi$ for all
    $\lambda\in\Lambda'$. This implies that
    $\lambda(F \llbracket \varphi \rrbracket (\alpha(x)))=1$.  Hence
    we obtain
    $(\bigcap_{\lambda\in\Lambda'} \lambda)(F \llbracket \varphi
    \rrbracket (\alpha(x))) = 1$ and finally
    $x\vDash [\bigcap_{\lambda\in\Lambda'} \lambda]\varphi $. \qed
  \end{description}	
\end{laterproof}

\noindent Note that this construction might again violate the cardinality
constraints of the logic. In particular, for the probabilistic case
(Example~\ref{ex:distr-order}) we have finite formulas, but countably
many modalities. However, if we assume that the set of labels $A$ is
finite and restrict the coefficients in the coalgebra to rational
numbers, every cone modality can be represented as the intersection of only finitely many minimal given modalities and so the encoding preserves finiteness.

\section{{\sc T-Beg}: A Generic Tool for Games and the Construction of
  Distinguishing Formulas}
\label{sec:tbeg}

\subsection{Overview}

A tool for playing bisimulation games is useful for teaching, for
illustrating examples in talks, for case studies and in general for
interaction with the user. There are already available tools,
providing visual feedback to help the user understand why two
states are (not) bisimilar, such as \textsc{The Bisimulation Game
  Game}\footnote{\texttt{\texttt{http://www.brics.dk/bisim/}}} or
\textsc{Bisimulation Games Tools}\footnote{\texttt{\texttt{https://www.jeroenkeiren.nl/blog/on-games-and-simulations/}}}
\cite{DBLP:journals/corr/EscrigKW16}. Both games are designed for
labelled transition systems and \cite{DBLP:journals/corr/EscrigKW16}
also covers branching bisimulation.

Our tool \tbeg{} goes beyond labelled transition system and allows to
treat coalgebras in general (under the restrictions that we impose),
that is, we exploit the categorical view to create a generic tool. As
shown earlier in Sections~\ref{sec:gen_strategies}
and~\ref{sec:gen_formula}, the coalgebraic game defined in
Definition~\ref{def:game} provides us with a generic algorithm to
compute the winning strategies and distinguishing formulas.

The user can either take on the role of the spoiler or of the
duplicator, playing on some coalgebra against the computer. The tool
computes the winning strategy (if any) and follows this winning
strategy if possible. We have also implemented the construction of
the distinguishing formula for two non-bisimilar states.

The genericity over the functor is in practice achieved as follows:
The user either selects an existing functor $F$ (e.g. the running examples of the paper), or implements his/her
own functor by providing the code of one class with nine methods
(explained below). Everything else, such as embedding the functor into
the game and the visualization are automatically handled by
\tbeg{}. In the case of weighted systems, \tbeg{} even handles the graphical representation.

Then, he/she enters or loads a coalgebra $\alpha: X \rightarrow FX$
(with $X$ finite), stored as \textit{csv} (comma separated value)
file. Now the user can switch to the game view and start the game by
choosing one of the two roles (spoiler or duplicator) and selecting a
pair of states $(x_0, x_1)$, based on the visual graph representation.
\begin{figure}[h]
	\centering
	\includegraphics[width=0.7\textwidth]{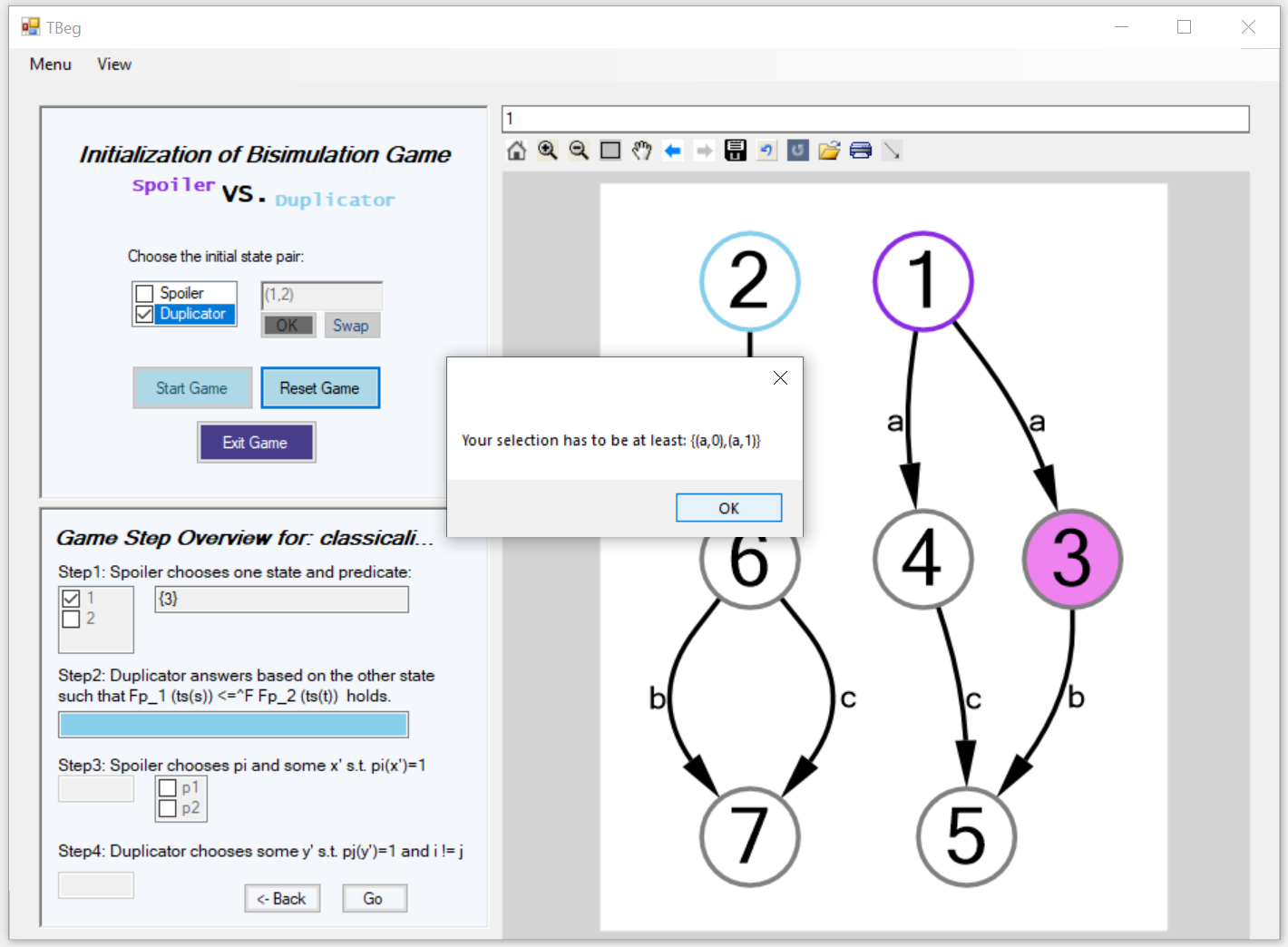}
	\caption{Screenshot of the graphical user interface with a game
		being played.}
	\label{fig:game-overview-tool}
\end{figure}
Next, the computer takes over the remaining role and the game starts:
In the game overview, the user is guided through the steps
by using two colors to indicate whether it is spoiler's
(\textcolor{violet} {violet}) or duplicator's (\textcolor{cyan}
{cyan}) turn (see Figure \ref{fig:game-overview-tool}).

In the case of two non-bisimular states, the tool will display a
distinguishing formula at the end of the game.

\subsection{Design}

We now give an overview over the design and the relevant methods
within the tool. We will also explain what has to be done in order to
integrate a new functor.

\tbeg{} is a Windows tool offering a complete graphical interface,
developed in Microsoft's Visual Studio using $C\#$, especially
\emph{Generics}. It uses a graph
library\footnote{\texttt{https://www.nuget.org/packages/Microsoft.Msagl.GraphViewerGDI}},
which in turn provides a \textit{GraphEditor} that allows for storing
graphs as \textit{MSAGL} files or as \textit{png} and \textit{jpg}
files.

The program is divided into five components: Model, View, Controller,
Game and Functor. We have chosen \emph{MVC (Model View Controller)} as
a modular pattern, so modules can be exchanged. Here we have several
$\mathit{Model\langle T\rangle}$ managed by the $\mathit{Controller}$,
where the functor in the sense of a \textit{Functor} class, which
always implements the \textit{Functor Interface}, is indicated by the
parameter $\langle T\rangle$.

While the tool supports more general functors, there is specific
support for functors $F$ with $F = V^{G(-)}$ where $V$ specifies a
semiring and $G$ preserves finite sets. That is, $F$
describes the branching type of a weighted transition system, where
for instance $G = A\times (-)+1$ (introducing finitely many labels and
termination). Coalgebras are of the form $X \to V^{GX}$ or -- via
currying -- of the form $X\times GX\to V$, which means that they can
be represented by $X\times GX$-matrices (matrices with index sets $X$,
$GX$). In the implementation $V$ is the generic data type of the
matrix entries. In the case of the powerset functor we simply have
$V = 2$ and $G = \mathit{Id}$.

If the branching type of the system can not simply be modelled as a matrix, there
is an optional field that can be used to specify the system, since
$\mathit{Model\langle T\rangle}$ calls the user-implemented method to
initialize the $F$-coalgebra instance.
The implementation of Algorithm~\ref{AlgorithmA_extended} can be found
in $\mathit{Game}\langle T,V\rangle$, representing the core of the
tool's architecture, whose correctness is only guaranteed for
  functors that meet our requirements, such as the functors used in
  the paper.

\subsubsection{Functor Interface}
As mentioned previously, the user has to provide nine methods in order
to implement the functor in the context of \tbeg{}: two are needed for the computation, two for
rendering the coalgebra as a graph, one for creating modal 
formulas, another two for loading and saving, and two more for
customizing the visual matrix representation.

We would like to emphasize here that the user is free to
formally implement the functor in the sense of the categorical
definition as long as the nine methods needed for the game are
provided. In particular, we do not need the application of the functor
to arrows since we only need to lift predicates
  $p:X \rightarrow 2$.

Within $\mathit{MyFunctor}$, which implements the interface
$\mathit{Functor\langle F, V \rangle}$, the user defines the data
structure $F$ for the branching type of the transition system (e.g., a
list or bit vector for the powerset functor, or the
corresponding function type in the case of the distribution
functor). Further, the user specifies the type $V$ that is needed to
define the entries of $X \times GX$ (e.g. a double value for a weight
or $0,1$ to indicate the existence of a transition).

\short{Here we focus on five methods which have to be provided,
  omitting the remaining four which are less central.} \full{Then the following nine methods has to be provided:}
\begin{description}
\item[$\mathit{Matrix\langle F, V\rangle InitMatrix(\dots)}$:] This
  method initializes the transition system with the string-based input
  of the user. The information about the states and the alphabet is
  provided via an input mask in the form of a matrix.
\item $\mathit{ bool \ CheckDuplicatorsConditionStep2( \dots) } $:
  given two states $x_0,x_1$ and two predicates $p_0,p_1$, this method
  checks whether
  \[ Fp_0 (\alpha(x_0)) \leq^F Fp_1 (\alpha(x_1)). \] This method is used
  when playing the game (in Step~$2$) and in the partition refinement
  algorithm (Algorithm~\ref{AlgorithmA_extended}) for the case
  $p_0=p_1$.

\item $\mathit{TSToGraph(\dots)}$: This method handles the
  implementation of the graph-based visualization of the transition
  system.
  For weighted systems the user can rely on the default
  implementation included within the \textit{Model}. In this case,
  arrows between states and their labels are generated automatically.
\item $\mathit{GraphToTS(\dots)}$: This method is used for the other
  direction, i.e.\ to derive the transition system from a directed
  graph given by $\mathit{Graph}$.

\item $\mathit{ string \ GetModalityToString(\dots)}$: This method is
  essential for the automatic generation of the modal logical formulas
  distinguishing two non-bisimilar states as described in
  Definition~\ref{def:constr_formu}. In each call, the
  cone modality that results from $ F\chi_P (\alpha (s)) $ with
  $ T (x_0, x_1) = (s, P) $ is converted into a string.
\full{\item $\mathit{SaveTransitionSystem(\dots)}$: In order to store a
  transition system in a \textit{csv} file.
\item $\mathit{LoadTransitionSystem(\dots)}$: In order to load a
  transition system from a \textit{csv} file.
\item $\mathit{GetRowHeadings(\dots)}$: \tbeg{} can visualize a transition system $\alpha: X \rightarrow FX$ as a $X \times GX$
  matrix within a \textit{DataGrid}. For this purpose, the user needs
  to specify how the \textit{RowHeaders} can be generated automatically.
\item $\mathit{ReturnRowCount(\dots)}$: This method returns
  the number of rows of the matrix representing the coalgebra.}
\end{description}
  
The implementation costs arising on the user side can be improved by employing a separate module
  that automatically generates functors (see
  \cite{coparFM19}). But it is not clear whether the lifting of the
  preorder can be obtained automatically. \full{Nevertheless, a combination
  of \cite{coparFM19} and \cite{DBLP:journals/corr/0001KM17aa}
  featuring \tbeg{} would result in a very powerful coalgebraic tool
  framework.}

\section{Conclusion and Discussion}
\label{sec:conclusion}

Our aim in this paper is to give concrete recipes for explaining
non-bisimilarity in a coalgebraic setting. This involves the
computation of the winning strategy of the spoiler in the bisimulation
game, based on a partition refinement algorithm, as well as the
generation of distinguishing formulas, following the ideas of
\cite{c:automatically-explaining-bisim}. Furthermore we have presented a
tool that implements this functionality in a generic way. Related
tools, as mentioned in \cite{DBLP:journals/corr/EscrigKW16}, are
limited to labelled transition systems and mainly focus on the spoiler
strategy instead of generating distinguishing formulas.

In the future we would like to combine our prototype implementation
with an efficient coalgebraic partition refinement algorithm, adapting
the ideas of Kanellakis/Smolka~\cite{Kanellakis1990CCSEF} or
Paige/Tarjan~\cite{DBLP:journals/siamcomp/PaigeT87} or using the
existing coalgebraic generalization~\cite{DorschEA17}, thus enabling
the efficient computation of winning strategies and distinguishing
formulae.

For the generation of distinguishing formulas, an option would be to
fix the modalities a priori and to use them in the game, similar to
the notion of $\lambda$-bisimulation
\cite{gs:sim-bisim-coalg-logic,km:bisim-games-logics-metric}. However,
there might be infinitely many modalities and the partition refinement
algorithm can not iterate over all of them. A possible solution would
be to find a way to check the conditions symbolically in order to
obtain suitable modalities.

Of course we are also interested in whether we can lift the extra
assumptions that were necessary in order to re-code modalities in
Section~\ref{sec:recode}. We also expect that the
  bisimulation game can be extended to polyadic predicate liftings.

An interesting further idea is to translate the coalgebra
into multi-neigh\-bour\-hood frames
\cite{h:monotonic-modal-logics,kw:monomodal-logics}, based on the
predicate liftings, and to derive a $\lambda$-bisimulation game as
in \cite{km:bisim-games-logics-metric,gs:sim-bisim-coalg-logic} from
there. (The $\lambda$-bisimulation game does not require weak
pullback preservation and extends the class of admissible functors,
but requires us to fix the modalities rather than generate them.)
One could go on and translate these multi-neighbourhood frames 
into Kripke frames, but this step unfortunately does not preserve
bisimilarity.

We also plan to study applications where we can exploit
the fact that the distinguishing formula witnesses
non-bisimilarity. For instance, we see interesting uses in the area of
differential privacy~\cite{cgpx:generalized-bisim-metrics}, for which
we would need to generalize the theory to a quantitative setting. That
is, we would like to construct distinguishing formulas in the setting
of quantitative coalgebraic logics, which characterize behavioural
distances.

\paragraph*{Acknowledgements} We would like to thank the reviewers for
the careful reading of the paper and their valuable
comments. Furthermore we thank Thorsten Wi{\ss}mann and Sebastian K\"{u}pper for inspiring
discussions on (efficient) coalgebraic partition refinement and
zippability.


%
%
%
\bibliographystyle{splncs04}
\bibliography{references}
%

\full{
\appendix

\section{Proofs}
\label{sec:proofs}

\doproofs

}

\end{document}